\documentclass[draft]{agujournal}
\draftfalse

\usepackage{graphicx}
\usepackage{amssymb}  
\usepackage{lineno}
\usepackage{amsmath}
\usepackage{siunitx}
\usepackage{tabularx}
\usepackage{algorithm2e}
\usepackage{amsmath}
\usepackage{algcompatible}
\usepackage{amsthm}
\usepackage{graphicx}
\usepackage{lscape}
\usepackage{verbatim}
\usepackage{color,soul}
\usepackage{xcolor}
\usepackage{subfigure}
\usepackage{tabularx,ragged2e,booktabs,caption,array,multirow,multicol}
\usepackage{hyperref}
\usepackage{csquotes}
\usepackage{mathtools} 
\usepackage{amsthm}  
\usepackage{listings}
\usepackage{amssymb}
\usepackage{latexsym}
\usepackage{epsfig}
\usepackage{float}
\usepackage{xspace}
\usepackage{algorithmwh}
\usepackage{float}
\usepackage[english]{babel}
\usepackage[utf8]{inputenc}
\usepackage{setspace}

\usepackage{subfigure}

\newcommand{\mb}[1]{\mathbf{#1}}
 \newcommand{\bs}[1]{\boldsymbol{#1}}
\usepackage{multirow}

\journalname{<enter journal name here>}

\begin{document}

%
%


\title{Multi-core parallel tempering Bayeslands for  basin and landscape  
evolution}

%
%


 \authors{Rohitash Chandra \affil{1,2}, R. Dietmar M\"uller\affil{1}, 
Danial Azam\affil{1}, Ratneel Deo \affil{1}, Nathaniel Butterworth  \affil{3}, Tristan Salles 
\affil{1}, Sally Cripps \affil{4}   }

  \affiliation{1}{EarthByte Group, School of Geosciences, University of 
Sydney, NSW 2006, Sydney, Australia}
  \affiliation{2}{Centre for Translational Data Science, University 
of Sydney, NSW 2006, Sydney, Australia}
\affiliation{3}{Sydney Informatics Hub, University 
of Sydney, NSW 2006, Sydney, Australia}
\affiliation{4}{School of Mathematics and Statistics, University 
of Sydney, NSW 2006, Sydney, Australia}
 




\correspondingauthor{Rohitash Chandra}{rohitash.chandra@sydney.edu.au}




\begin{keypoints}
\item 
Landscape evolution 
\item Bayesian inference
\item Parallel tempering 
\item Badlands 
\item Bayeslands 
\end{keypoints}

%
%


\begin{abstract}

 The Bayesian paradigm is becoming an increasingly popular framework for  estimation and   uncertainty quantification of  unknown parameters in geo-physical inversion problems.  Badlands is a basin and landscape  evolution forward model for simulating topography evolution at a large range of spatial and time 
scales. Our previous work  presented Bayeslands that used  the Bayesian paradigm  to make inference for  unknown parameters in the  Badlands model using  Markov chain Monte Carlo (MCMC) sampling. Bayeslands faced  challenges in convergence due to multi-modal posterior distributions in the selected parameters of Badlands.   Parallel tempering     is an advanced  MCMC method  suited for irregular and  multi-modal posterior  distributions. In this paper, we extend Bayeslands using  parallel tempering (PT-Bayeslands)    \textcolor{black}{with}  high performance   computing to address previous limitations in parameter space exploration in the context of the  computationally expensive Badlands model.  Our results show that PT-Bayeslands not only reduces the computation time, but also \textcolor{black}{ provides an improvement of the sampling} for multi-modal posterior distributions. This provides an improvement over Bayeslands which  used  single chain MCMC that face difficulties in convergence and can lead to misleading inference. This motivates its usage  in large-scale  basin and landscape evolution models.

\end{abstract}

%
%

\section{Introduction}
\label{S:1}
  

Understanding landscape evolution, and the associated accumulation of sediments in basins, is limited by increasingly sparse data back through geological time.  Recent developments in landscape evolution models (LEMs) \citep{coulthard2001landscape}, such as the basin and landscape dynamics (Badlands) model feature  responses to surface uplift and subsidence over a large range of spatial scales, and track sediments from source to sink \citep{salles2016badlands,salles2017influence,salles2018pybadlands}.  Badlands also has  the capability to create synthetic basin stratigraphies.  LEMs depend on uncertain initial and boundary conditions \citep{scott1990finite}; such as the initial topography, sea level, precipitation, and rock lithology and erodibility \citep{godard2006numerical,rejman1998spatial}. Quantifying uncertainty surrounding these initial conditions and how the topography changes over time is an incredibly  difficult task.

 To make inference about  the parameters that govern landscape evolution, we need to constrain the  model  by fusing information from various sources in a probabilistic model. The sources of information include \textcolor{black}{ underlying equations that govern  LEMs to   represent  geophysical processes},  observed data  such as present day topography,   stratigraphy in sedimentary basins, and  previous research which could be denoted as expert knowledge \citep{salles2018pybadlands}.


The Bayesian framework provides a logically consistent mechanism for fusing information from various sources  to provide meaningful inference  for unknown parameters in models.  The prior distribution is a mechanism to incorporate information from previous research and  expert opinion, and the likelihood is a mechanism to incorporate information from data to sample from the posterior distribution that represents the unknown parameter\citep{tarantola2006popper}.   The ability to fuse information from many sources in a principled fashion has made Bayesian inference an increasingly popular choice for the estimation and uncertainty quantification of parameters in complex models
\citep{robert2011short,mosegaard1991simulated,rocca2009evolutionary,sen2013global,gallagher2009markov}. \textcolor{black}{More specifically, there is some work that employs Bayesian inversion methodologies for landscape evolution based forward models for specific applications. Initial work was done by 
\citep{roberts2010estimating} who used Monte Carlo inverse modeling of   river profiles to estmate uplift rate histories using African examples.  
\citep{fox2014rock}  reviewed uplift and erosion rate history of the Bergell intrusion from the inversion of low temperature thermochronometric data  and   
\citep{goren2014tectonics} focused on distinguishing   tectonics from fluvial topography using formal linear inversion with  applications to the Inyo Mountains, California.  \citep{fox2015abrupt} investigated    the rate of Andean Plateau uplift using  reversible jump MCMC inversion of river profiles.    }

 Although Bayesian  inversion has become popular in geophysics in the past few decades \citep{grandis1999bayesian,malinverno2002parsimonious,mosegaard1995monte,sambridge2002monte,
sambridge1999geophysical},  estimating  the posterior distribution is often nontrivial. The posterior distributions can be  multimodal, exhibit discontinuities and is usually not available in closed form; hence, Markov chain Monte Carlo (MCMC) sampling \textcolor{black}{methods} are used  to estimate it and crafting proposal distributions to explore these distributions  is very difficult 
\citep{gallagher2009markov}.

The Metropolis-Hastings algorithm  \citep{hastings1970monte,metropolis1953equation,chib1995understanding} is a popular MCMC  method   to obtain iterations from a distribution that cannot be sampled directly.  In the Metropolis-Hastings algorithm, the  current state  in the Markov chain, is moved to a new state via a transition kernel.  This transition kernel consists of a proposal distribution and an acceptance probability. Given some regularity conditions, this acceptance probability  is constructed so that the Markov chain will converge \textcolor{black}{to} the posterior (stationary)  distribution, irrespective  of the proposal distribution. However, the proposal distribution does effect the rate at which the chain converges. A poorly chosen proposal distribution can result  in an acceptance probability close to zero and the chain will not converge in a timely fashion. Information about the gradient of the posterior may help with the construction of better proposal distributions  \citep{neal2011mcmc,hoffman2014no,girolami2011riemann}; however,  in  geophysical forward models, gradients are   often  unavailable or too computationally expensive to  obtain. In situations such as these, \textcolor{black}{ parallel tempering  Markov Chain Monte Carlo}   (PT-MCMC) \citep{marinari1992,geyer1995annealing}  offers the best chance of convergence. 

PT-MCMC is well suited for  exploring multi-modal distributions by running replicas of the Markov chains in parallel. In PT-MCMC, the replicas converge to  different stationary distributions which effectively smooth out the local modes that exist in the posterior. The parameters in replicas  are allowed to swap with each other, ensuring that the target chain will contain draws from other local maxima.
In addition, parallel computing based implementation of PT-MCMC can  reduce  the computational burden associated with complex forward models such as Badlands  \citep{zhang2007toward,vrugt2006application,mills1992implementing}. The potential for PT-MCMC in
 geoscience has been demonstrated    for  complex 
 multimodal   problems     
 \citep{sambridge2013parallel, sen1996bayesian,maraschini2010monte,sen2013global, Scalzo2018GMD}.  Although recent implementations of PT-MCMC with gradient-based proposals have shown  promising performance   \citep{CHANDRA_NC2019},  \textcolor{black}{they} are not feasible  for the Badlands model because gradients are unavailable from the model.
 
 Although MCMC methods with  random-walk  proposals  are feasible for models that do not provide gradients, they are inefficient and likely to get  stuck in local modes \citep{neal1996sampling}. In our previous work  \citep{CHANDRA2019}, we presented the Bayeslands framework that used single chain   random-walk Metropolis Hastings   as a proposal distribution to obtain draws from the posterior distributions of interest. \textcolor{black}{ The method did not use parallel computing, and hence referred to as single chain Bayeslands (SC-Bayeslands), hereafter}.   SC-Bayeslands demonstrated that even in low dimensional settings, the posterior surfaces of parameters  exhibited highly irregular
features, such as \textcolor{black}{multimodality} and discontinuities, making sampling  difficult.

  In this paper, we present a multi-core parallel tempering  Bayeslands (PT-Bayeslands) that features uncertainty quantification and estimation of selected parameters for basin and landscape evolution. We investigate a number of issues such as sampling multi-modal posterior distributions, proposal distributions,  computational time   and prediction quality of Badlands. We apply our  technique to simulated data  and real-world topographies.


\section{Background and Related Work}
 
\label{sec_background}
  
\subsection{Badlands}

Over the last decades, many numerical models have been proposed to simulate how 
the Earth surface has evolved over geological time scales in response to 
different driving forces such as tectonics or climatic variability 
\citep{Whipple2002,Tucker10,Salles15,Campforts2017,Adams2017}. These models 
combine empirical data and conceptual methods into a set of mathematical 
equations that can be used to reconstruct landscape evolution and associated 
sediment fluxes \citep{Howard1994,Hobley2011}. They are currently used in many 
research fields such as hydrology, soil erosion, hillslope stability and general landscape studies.


We use Badlands 
\citep{salles2016badlands,salles2016b,salles2018a} to
simulate regional to continental sediment deposition and associated 
sedimentary basin architecture \citep{salles2017influence,salles2018b}.  In its 
\textcolor{black}{basic} formulation, the Earth surface elevation change, denoted by $\frac{\partial z}{\partial t}$; where $z$ is the elevation and $t$ refers to time in years   related to the 
interaction of three types of processes: the tectonic uplift rate ($U$), the incision rate by rivers ($I$) and the hillslope processes ($D$). 
\begin{equation}
\frac{\partial z}{\partial t}=U-I+D
\label{eq:lem}
\end{equation}

In this study, fluvial incision rates and predicted sediment transport in rivers are solved using the stream-power law (SPL) \citep{stock1999geologic,HAREL2016184}. SPL relates the erosion rate  $I$ to  the product of mean annual net precipitation rate ($\bar{P}$), drainage area ($A$),  and local river gradient ($S$) and takes the form:
\begin{equation}
I= k (\bar{P}A)^\gamma S^\lambda
\label{eq:eqn_erosion}
\end{equation}
where, $k$ is an erodibility coefficient that depends on lithology and climate, while $\gamma$ ($<2$) and $\lambda$ ($<4$) are positive exponents \citep{Chen14} that mostly depend on catchment hydrology and the nature of the dominant erosional mechanism \citep{Whipple2002}. Despite its simplicity, Equation~\ref{eq:eqn_erosion} reproduces many of the characteristic features of mountainous landscapes, where detachment-limited erosion regime dominates \citep{Tucker10}. $\kappa$ varies by several orders of magnitude not only based on lithology, climate, sedimentary flux or river channel width but also with the chosen values of $\gamma$ and $\lambda$.

\noindent In addition to overland flow, semi-continuous processes of soil displacement are accounted for using a linear diffusion law commonly referred to as soil creep \citep{Tucker10}:
 \begin{equation}
 D=\delta \nabla^2 z
 \label{eq:eqn_soil}
\end{equation}
where $\delta$ is the diffusion coefficient. This transport rate depends linearly on topographic gradient and encapsulates in a simple formulation the processes operating on superficial sedimentary layers.

 \subsection{SC-Bayeslands}

 \textcolor{black}{The posterior distributions of parameters in geophysical inversions problems   are notoriously difficult to sample \citep{sambridge1999geophysical}. They are high dimensional, multimodal and sometimes exhibit discontinuities.
 SC-Bayeslands provides a framework for traversing the posterior distribution of selected parameters of   the Badlands model. In our previous work \citep{CHANDRA2019}, we found that SC-Bayeslands had difficulty in convergence due to multimodal posterior distributions. In the experiments, we used two problems, namely the synthetic crater (Cr) and the continental margin (CM) problem to demonstrate effectiveness of Bayeslands. We considered   estimation    of \textit{precipitation} and \textit{erodibility}, which are  the key parameters  in the Badlands model.}

\begin{figure}[htbp!]
  \begin{center}  
   \includegraphics[width=100mm]{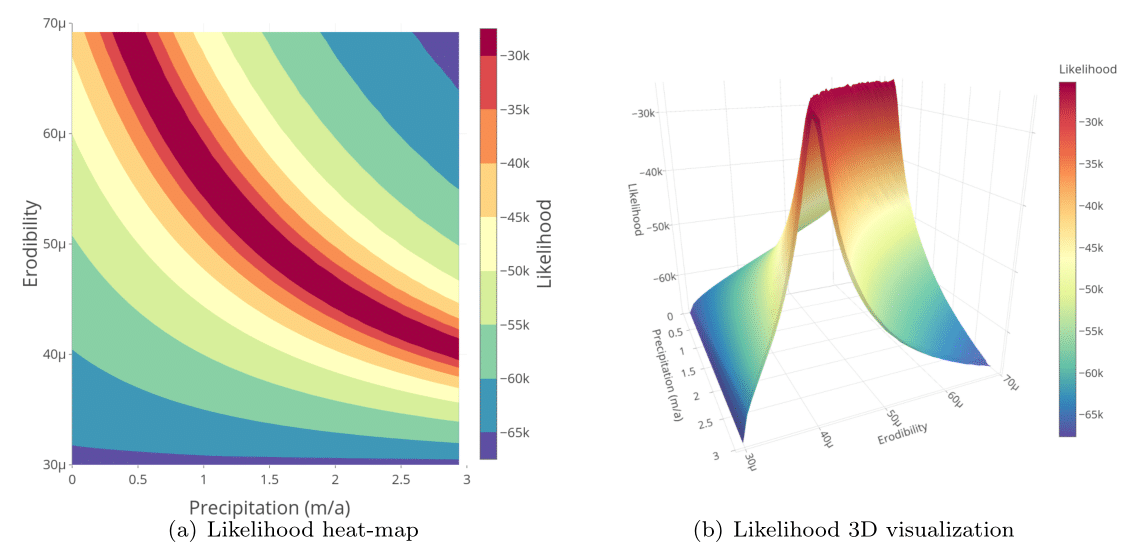} \\
    \caption{Panels~(a)~and~(b) are a heat map and a surface plot of the log posterior  surface of the Cr  as function of   precipitation, $\rho, $ and erodibility, $\epsilon$ \citep{CHANDRA2019}.}
 \label{fig:lks_craterx}
  \end{center}
\end{figure}

\begin{figure}[htbp!]
  \begin{center}  
   \includegraphics[width=100mm]{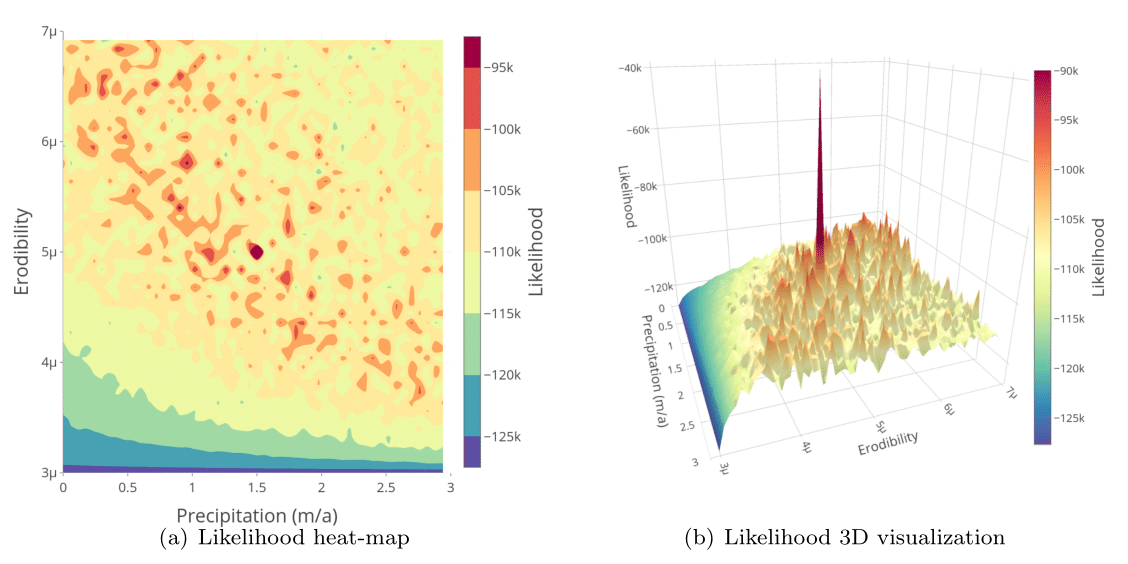} \\
    \caption{Panels~(a)~and~(b) are a heat map and a surface plot of the log posterior  surface of the CM problem as function of   precipitation, $\rho, $ and erodibility, $\epsilon$. The true values have a clear peak in the log-posterior (precipitation=1.5\,m/a, erodibility=$5e^{-6}$), however many sub-optimal  peaks (marked in red) for this parameter pair exist \citep{CHANDRA2019}. }
 \label{fig:lks_cm}
  \end{center}
\end{figure}

\textcolor{black}{Figures ~\ref{fig:lks_craterx} and  \ref{fig:lks_cm} show examples of  the posterior distributions for   precipitation   and   erodibility, while fixing all other parameters, for the Cr and CM problems, respectively.  The Cr log-posterior surface in Figure \ref{fig:lks_craterx} is  smooth but has a clearly defined ridge, which SC-Bayeslands has difficulty exploring.  We note that our previous work  \citep{CHANDRA2019} showed that  different combinations  of precipitation  and erodibility  gave rise to visually indistinguishable topography for both the Cr and CM problem.  Figure~\ref{fig:lks_cm} has a clear global maximum but also several local maxima. Here,  unless  the  chain has a starting value near this global maximum, it is unlikely that it would be found by SC-Bayeslands. }
  
 \textcolor{black}{Furthermore, the problems (Cr and CM) used in SC-Bayeslands  took computational time of a few seconds and hence it was possible to take thousands of samples for convergence. There is impracticability in applying the SC-Bayeslands for landscape evolution problems that have a computational time of several minutes to few hours, hence parallel computing is needed. }

 \subsection{Parallel tempering MCMC }

PT-MCMC  features an ensemble of  replicas (samplers) that are executed  in parallel with the ability to explore multi-modal posterior distributions \citep{geyer1995annealing}.   Parallel tempering carries out an exchange of parameters in neighbouring replicas during sampling that is helpful in escaping local minima.  In other words, the Markov chains in the ensemble of   replicas have stationary distributions which 
are equal to (up to a proportionality constant) $p(\theta|\mb D)^{\beta}$, where 
$\beta\in[0,1]$ and is known as the {\it temperature}  of the replica. A replica which has a temperature of $\beta=0$ has a stationary distribution which is uniform while one which has a temperature $\beta=1$ corresponds to a stationary distribution which  is the 
 posterior  $p(\theta|\mb D)$. This means that the replica  where $\beta<<1$ have  local maxima which are less separated, and therefore, replicas with smaller values of  $\beta$ are less likely to get stuck in a local minima and thus explore a 
larger region.  The replicas with higher values of $\beta$ 
typically explore local regions.  The choice of the temperature ladder and number of replicas governs the computational efficiency of the PT-MCMC  and is the subject of much research. For instance,   \citep{kone2005selection,patriksson2008temperature}  describe an efficient method of finding the temperature ladder, and \citep{miasojedow2013adaptive}   proposed a method for attentively adapting the temperature ladder based on the performance.

 The other feature of parallel tempering is their feasibility of implementation in multi-core or parallel computing architectures. In multi-core implementation, factors such as inter-process communications need to be considered during the exchange  between the neighboring replicas \citep{lamport1986interprocess}. Effective communication strategies between the processes that execute the respective  replicas needs to be made in order to reduce the overhead of processes waiting for others. Hence,  a decentralized  
implementation of parallel tempering  is presented that  eliminates global 
synchronization and
reduces the overhead caused by inter-process communication in exchange of 
solutions between the chains that run in parallel cores \citep{LI2009269}. 
Parallel tempering has also been implemented  in a 
distributed volunteer computing network  where computers belonging to the 
general public are used with help of multi-threading and graphic processing 
units (GPUs) \citep{karimi2011high}. The implementation with Field Programmable Gate Arrays (FPGAs) that has massive 
parallelism capabilities resulted in much better performance than multi-core 
and GPU implementations \citep{mingas2017particle}. 
In terms of applications, other studies have efficiently implemented parallel tempering via 
multi-core architectures for  exploration of Earth's resources  
\citep{reid2013bayesian}.


\section{Methodology}

\subsection{Synthetic topography data}
\label{sec:syn_topo}


\textcolor{black}{
Badlands is a forward stratigraphic model that takes at time $t=0$, an initial topography denoted by $\mb D_0=(D_{s_10},\ldots,D_{s_n0})$; where $D_{s_i0}$ is the elevation at site $s_i$ at time $t=0$, for $i=1,\ldots,n$, given    $n$ years. The parameters in Badlands model are denoted by $\theta$ that are used to produce a series of topographies, $\mb D_0,\mb D_1,\ldots,\mb D_T$. } We assume that the   final topography   $\mb D_T$, is the only topography to compare with the ground-truth topography. Therefore, the task of making inference about the  landscape evolution over time is very difficult.

The four key input parameters  represented by  $\theta$  include  precipitation ($\rho$), rock erodibility  ($k$), and  the contribution of rain  and slope on erodibility denoted by m-value ($\mu$) and n-value ($\nu$), respectively. Additional   parameters depending on the nature of the landscape evolution problem include  surface ($\gamma$) and  marine ($\beta$)  which govern   coastal landscapes. In problems where mountain building processes are involved, the tectonic uplift ($u$) parameter is utilized.   The relationship between these variables and landscape evolution is given by  Equations~ \ref{eq:lem}- 
 \ref{eq:eqn_soil}.  The function that maps the initial topography and these parameters to the final  topography (time $t=T$)  is denoted by  $\mb f_T(\mb D_0,\bs\theta)=\left(f_{T,s_1}(\mb D_0,\bs\theta),\ldots,f_{T,s_n}(\mb D_0,\bs\theta)\right)$, where $f_{T,s_i}$ is the elevation at time $T$ for site $s_i$, for $i=1,\ldots,n$. 
 
 Different landscape evolution trajectories    could lead to the same final topography, and to constrain the number of possible trajectories requires additional sources of information.  One source of information is the history of  sediment erosion/deposition at various locations.  We denote the sediment erosion/deposition at time $t$ by 
 $\mb z_t=\left(z_{s_1t},\ldots,z_{s_jt}\right)$; where $z_{s_j,t}$ is the sediment erosion/deposition at site $s_j$ for $j=1,\ldots,J$. We define another function that maps $\mb D_0$  and $\bs\theta$  to sediment erosion/deposition by 
 $\mb g_{t}(\mb D_0,\bs\theta)=\left(g_{s_1t}(\mb D_0,\bs\theta),\ldots,g_{s_mt}(\mb D_0,\bs\theta)\right)$, for $t=1,\ldots,T$. 
  
\begin{figure}[htbp!]
  \begin{center}  
   \includegraphics[width=100mm]{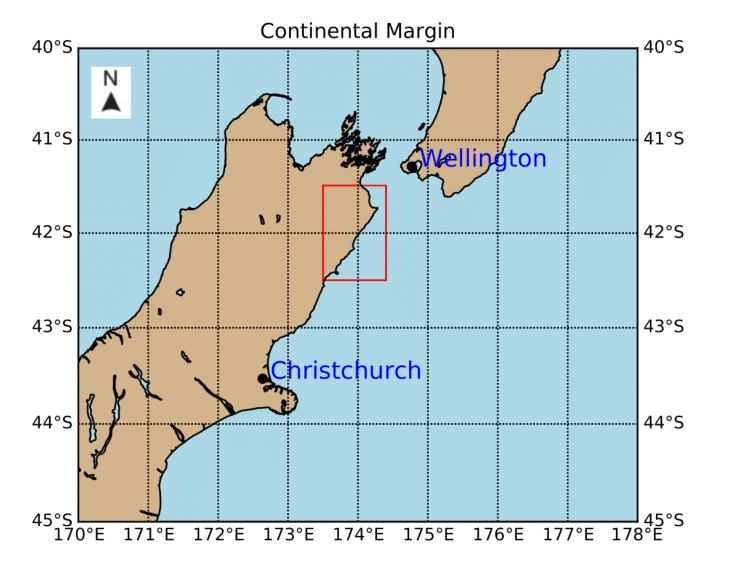} \\
    \caption{CM problem  selected from South Island of New Zealand, outlined  by the red rectangle. Note: x-axis represents longitude, while y-axis represents latitude.}
 \label{fig:cm-map}
  \end{center}
\end{figure}

\begin{figure}[htbp!]
  \begin{center}  
   \includegraphics[width=100mm]{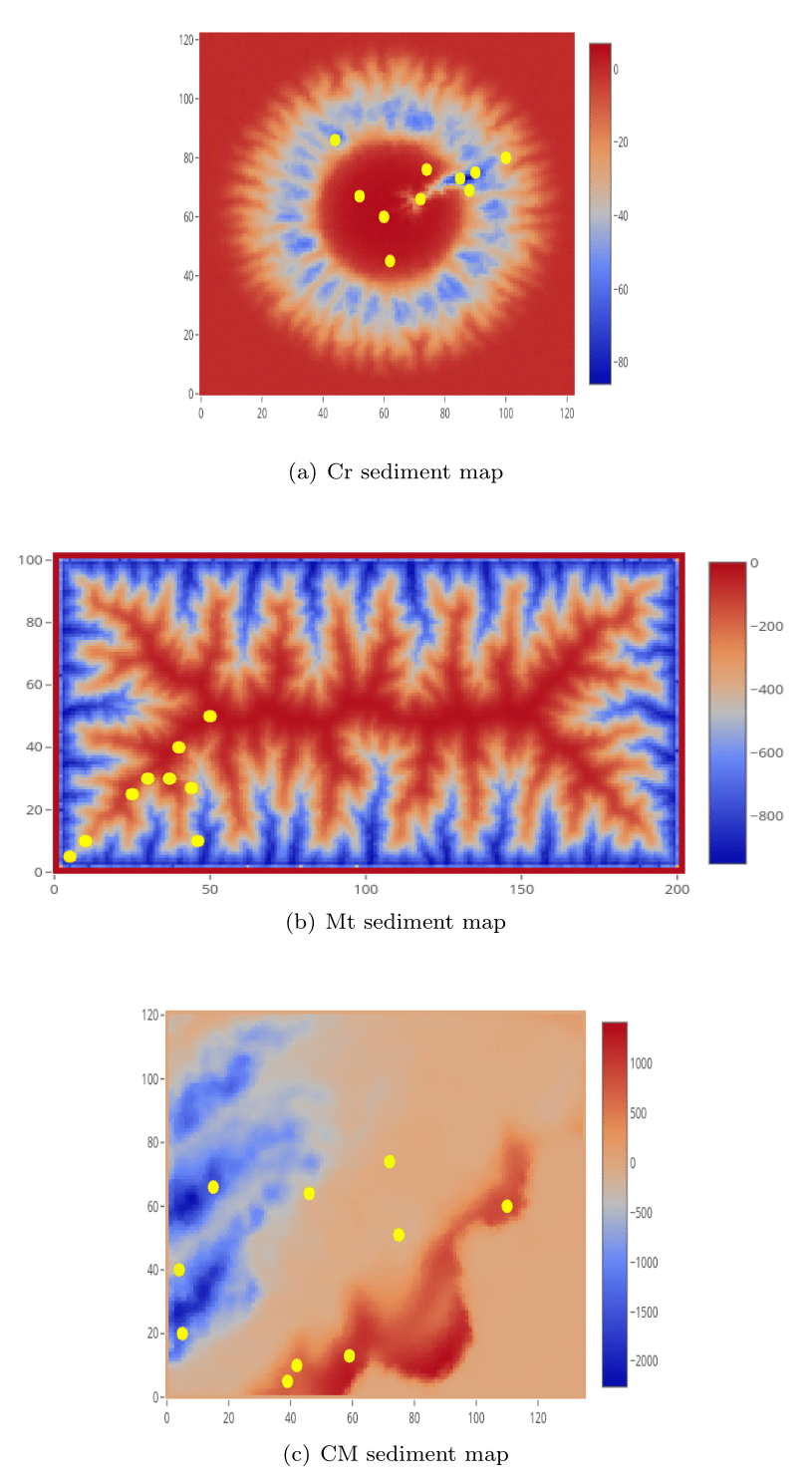} \\
    
    \caption{Initial (first column) and final or ground-truth (second column) topographies. Panels (a)~and~(b) refer to the Synthetic-Crater landscape for $t=0$ and $t=50,000$ years respectively. Panels (c)~and~(d) refer to the  Synthetic-Mountain for $t=0$ and $t=1,000,000$ years respectively, while Panels (e)~and~(f) refer to the Continental-Margin landscape for $t=0$ and $t=1,000,000$ years respectively.}
 \label{fig:craterdata}
  \end{center}
\end{figure}

\textcolor{black}{In order to test the proposed methodology, we first use Badlands to create synthetic ground-truth topography data for  two synthetic  and a real-world landscape evolution problem. We use the present day topography as the initial model topography. We use Badlands to model landscape and basin evolution given values for selected parameters (Table 2) and a specified time $T_{max}$ (Table 1) given by  number of years to simulate the ground-truth data for topography and sediment erosion/deposition. We refer to the selected  problems as   Synthetic-Crater (Cr),   Synthetic-Mountain (Mt)  and Continental Margin (CM) problems. The first two problems have a synthetic initial topography (hence the name) while the third uses a real landscape, in the North-Eastern region of the South Island in New Zealand  as its initial topography (Figure \ref{fig:cm-map}).} In all the problems, the  Badlands model takes selected parameter values   together with the initial topographies to produce final topographies.  The initial and final topographies for all three problems are shown in Figure \ref{fig:craterdata}. \textcolor{black}{ We note that Cr and CM  has been adapted from  examples of the  Badlands model \footnote{\url{https://github.com/badlands-model/pyBadlands/tree/master/Examples}},  and also used in our previous work \citep{CHANDRA2019}.}

These three problems were chosen to highlight the impact of different parameters on the final topography.  In the Cr problem the parameters are precipitation $\rho$, rock erodibility $k$,  the contribution of rain and slope on erodibility $\gamma$ and  $\lambda$ respectively, so that $\bs\theta_{Cr}=(\rho,k,\mu,\nu)$. The Mt problem starts off with a much simpler initial topography, but features the impact of tectonic uplift, $u$ as well as all the four inputs  in Cr problem, so that $\bs\theta_{Mt}=(\rho,k,\mu,\nu,u)$.  The CM  problem  does not feature tectonic uplift; however, since the location covers  coastal region, it includes parameters which model the impact of the surface and marine environment  on landscape evolution, so that $\bs\theta_{CM}=(\rho,k,\mu,\nu,\beta,\lambda)$. 

 \textcolor{black}{The parameter $\mu$ is particularly important in that it describes the sensitivity of a tectonically active mountain belt to changes in precipitation or tectonic accretion. It also defines how incision rates will change as the discharge becomes flashier  \cite{gasparini2011generalized}.}

\begin{figure}[htbp!]
  \begin{center}  
   \includegraphics[width=140mm]{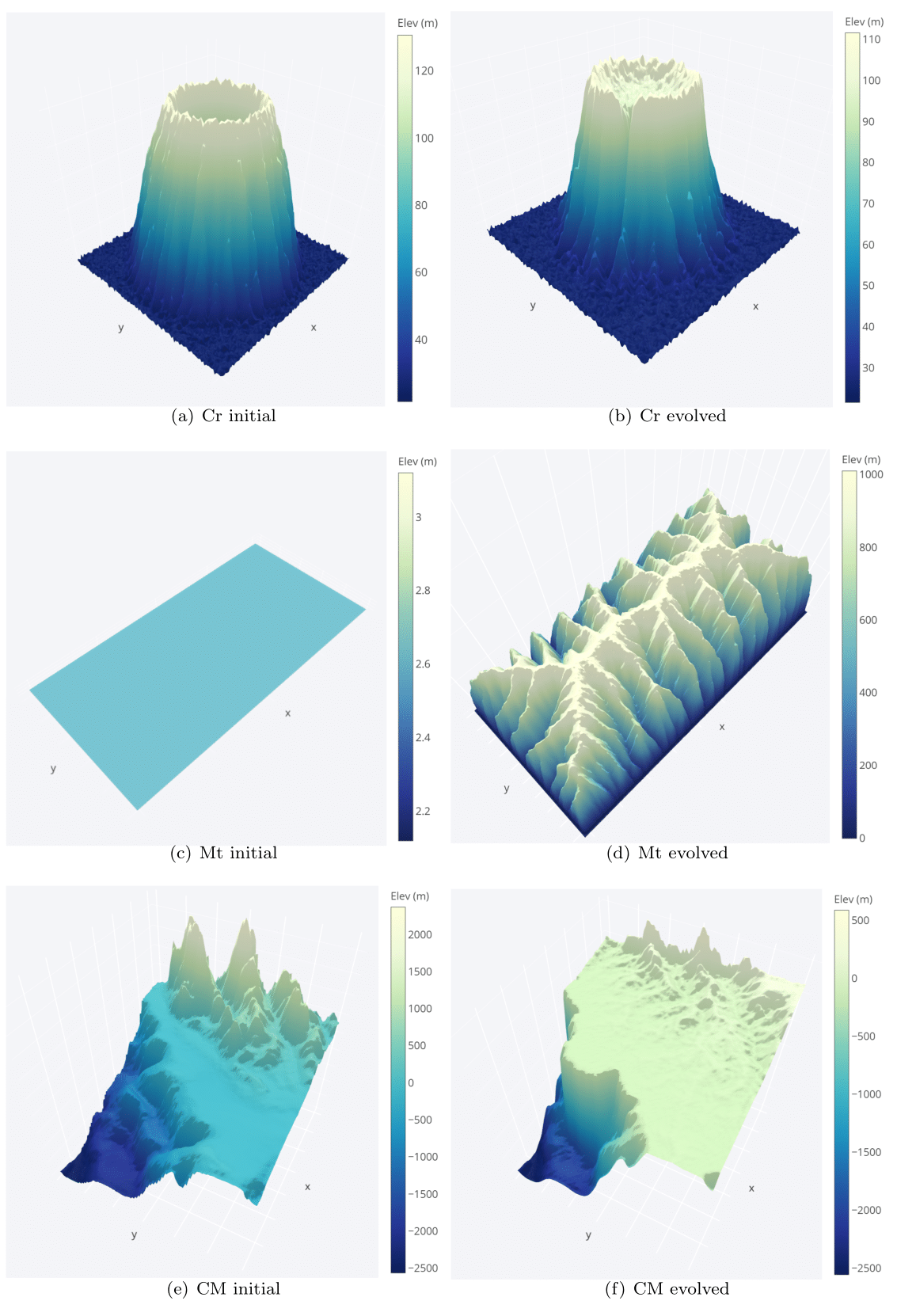} \\

    \caption{ Map view showing evolved model sediment erosion  and deposition for each scenario. The selected locations   for likelihood evaluation are overlain as yellow dots. The color bars indicate elevation in meters while the x and y axis give the spatial coordinates.  Note that \textcolor{black}{positive} values indicate deposition and  the negative values indicate erosion.  }

 \label{fig:sediments}
  \end{center}
\end{figure}

 Table \ref{tab:problem} contains descriptions of the problem such as the evolution time, the area of the landscape given by its width and length, the amount of time taken to run the Badlands model, and the resolution factor  (Res. factor) which  indicates the distance in  kilometres  (km) between two neighbouring points (pts) along x or y-axis of the topography grid.  Table~\ref{tab:truevalues} give details of the initial conditions used in the simulations for the various problem. Each of these problems features  erosion/deposition of sediment  over time.   Figure 
\ref{fig:sediments} shows the change in sediment erosion/deposition  at the final stage of evolution given by the Badlands model. Note that   \textcolor{black}{the positive} values indicate deposition and  the negative values indicate erosion.  The yellow dots indicate the locations for  the ground-truth data for  sediment erosion/deposition history.  
Hence, the likelihood function  given in the following  subsection takes both the landscape topography and erosion-deposition ground-truth into account.   

\begin{table*}[htbp!]
\smaller
\centering
 \begin{tabular}{ l c c c c  c } 
 \hline 
Topography & $T_{max}$ (years)&  Length [km, pts] &  Width [km, pts]   & Res. factor & 
Run-time (s)  \\  
 \hline   
 Cr &  50 000 & [0.24, 123]  & [0.24, 123]& 0.002 &  2.0\\  
 Mt  &  1 000 000 &  [80,202] &  [40,102]& 1.000  & 10.0 \\
 CM & 1 000 000  &  [136.0, 136] &  [123.0, 123]& 1.000 &  7.5\\ 
 \hline
 \end{tabular}
 
\caption{Landscape evolution problems  where the run-time 
represents the approximate length of time for one sample (simulation by Badlands) to run. The resolution factor (Res. factor)  indicates the distance in  kilometres  (km) between two neighbouring points (pts) along x or y-axis of the topography grid. $T_{max}$ denotes the  maximum evolution time in years. } 

 \label{tab:problem} 
\end{table*}

\begin{table*}[htbp!]
\smaller
\centering
 \begin{tabular}{ l c c c c  c  c c } 
 \hline 
Topography & $\rho$ (m/yr)&  $k$ & $\nu$  & $\mu$ & $\beta$ & 
$\gamma$ & $ u$   \\
 \hline  
Cr &  1.5 &  5.0-e05   &1.0  & 0.5 & - & - & - \\ 
 Mt & 1.5  & 5.0-e06  & 1.0  & 0.5 &  -  & - & 1.0 \\  
 CM  & 1.5  & 5.0-e06  & 1.0  & 0.5 &  0.5  & 0.8 & - \\

 \hline

 \end{tabular}
 
\caption{Selected values of parameters used to generate synthetic ground-truth topographies.  
The  parameters include  precipitation ($\rho$) in meters per year (m/yr), erodibility ($k$), m-value ($\mu$), n-value ($\nu$), marine ($\beta$), surface ($\gamma$) and uplift ($u$) in millimeters per year (mm/yr) 
}
 \label{tab:problem}  
 \label{tab:truevalues} 
\end{table*}

\subsection{Model and Priors for Parallel Tempering Bayeslands}

    
    

\textcolor{black}{We assume that topography at time $t$, and location $s_i$ i.e. $D_{s_it}|\bs \theta,\mb D_0$, has normal distribution with an expected value equal to the  Badlands  model, given $\bs\theta$ and a variance of $\tau^2$ so that
\begin{equation}
D_{s_i,t}=f_{s_i,t}(\bs \theta)+e_{s_i,t}\;\mbox{with}\; 
e_{s_i,t}\sim N (0,\tau^2)
\label{eqn:topo_sim}
\end{equation}
for $t=0,1,\ldots,T$ and $i=1,\ldots,n$; where, $n$ refers to number of observations and $T$ is the maximum simulation time in years. Note that we assume that the observations are independent because the correlation structure of the topography is embedded in the Badlands model.   }

\textcolor{black}{Although the  choice of a normal distribution was made for convenience,  we note that there could have been several other choices.  For example, we could have assumed that the log of the topography had a normal distribution  to allow for negative predicted values which the model given by Equation~\ref{eqn:topo_sim} is capable of producing. We could also assume that the errors were distributed according to a  \textit{t-distribution}, or we could model the distribution of errors non-parameterically.}    

In the model given by Equation~\ref{eqn:topo_sim}, we assume an inverse gamma (IG) prior  $\tau^2\sim IG(\nu/2,2/\nu)$. We integrate it  so that  the likeihood for the topography at time $t=T$ is 
$L_l(\bs\theta)$ is 

\begin{equation}
 L_{l}(\bs\theta)\propto \prod_{i=1}^n \left(1+\frac{(D_{s_i,T}-f_{s_i,T}(\bs\theta))^2}{\nu}\right)^{-\frac{\nu+1}{2}}
\end{equation}
where, the subscript $l$, in $L_l(\bs\theta)$  denotes that it is the landscape likelihood to distinguish it from a sediment likelihood.


\begin{table*}[htbp!]
\smaller
\centering
 \begin{tabular}{ l c c c c  c  c c } 
 \hline 
Topography & $\rho$ (m/yr)&  $k$ & $\nu$  & $\mu$ & $\beta$ & 
$\gamma$ & $ \lambda$   \\  
 \hline  
Cr & [0,3.0 ]  & [3.0-e05, 7.0-e05] & [0, 2.0]  & [0, 2.0] & -  & -  & 
\\   

Mt & [0,3.0 ]  & [3.0-e06, 7.0-e06] & [0, 2.0] & [0, 2.0]  & -  & - & 
[0.1, 5.0] \\ 

CM & [0,3.0 ]  & [3.0-e06, 7.0-e06] & [0, 2.0]  & [0, 2.0] & [0.3, 0.7]  & 
[0.6, 1.0] & - \\ 
 
 \hline

 \end{tabular} 
 
\caption{Prior distributions of Badlands parameters  for precipitation ($\rho$) in meters per year (m/yr), erodibility ($k$), m-value ($\mu$), n-value ($\nu$), marine ($\beta$), surface ($\gamma$) and uplift ($ \lambda$) in millimeters per year (mm/yr) }
 \label{tab:priors} 
\end{table*}

 As noted in Section 3.1, the history of sedimentary deposits is available at some locations and we use this information to further constrain the number of possible landscape evolution trajectories.  Again, we assume that observed sediment erosion/deposition
values at time $t$,  $\mb z_t$, are a function of the Badlands   model, given  
$\bs\theta$ plus some Gaussian noise
\begin{equation}
z_{s_j,t} =g_{s_j,t}(\bs \theta)+\eta_{s_j,t}\;\mbox{with}\; \eta_{s_jt}\sim 
(0,\chi^2)
\label{eqn:sed_sim}\end{equation}
Analogous to the likelihood function for the topography,
the sediment likelihood  $L_s(\bs\theta)$, after integrating out $\chi^2$  is 
\begin{equation}
 L_{s}(\bs\theta)\propto \prod_{t=1}^T\prod_{j=1}^J \left(1+\frac{(z_{s_j,t}-g_{s_j,t}(\bs\theta))^2}{\nu}\right)^{-\frac{\nu+1}{2}}
\end{equation}

 We assume that the elevation observations are independent of the sediment observations, so that
 \begin{equation}
 L(\bs\theta)=L_s(\bs\theta)\times L_l(\bs\theta)
 \end{equation}
 While this assumption may not hold for the simple Cr example, it is not an  unreasonable assumption for the CM problem.  The complicated topology of the CM problem, together with its coastal location, means that the sediment deposition is dispersed rather than stationary as in the Cr problem.

We use uniform priors for each of the parameters with the lower and upper limits given in Table  
\ref{tab:priors}    which reflects on the  actual environmental and geological  processes. For instance, logically, the values of precipitation and erodibility cannot be below zero. The upper limit is determined by trial experiments and also literature review \citep{
salinger1980new,ummenhofer2007interannual}, in order to have realistic assumption for the given problems. For instance, we reviewed the precipitation for South Island in New Zealand over the last few decades and gathered that the upper limit of 3 meters per year (m/a) would be  a realistic assumption.

  \subsection{Sampling Scheme}
  \subsubsection{Within Replica Sampling:   proposal distributions}
 \textcolor{black}{PT-Bayeslands  employs a random-walk sampler for each replica   which is executed  as a separate  process. The random-walk proposal distribution, denoted generically by $q(\bs\theta|\bs\theta^c)$ is a multivariate normal distribution   with mean vector $\bs\mu$ and covariance matrix $\Sigma$. The mean vector $\bs\mu$ is set to the current value in the chain, $\bs\theta^c$. We investigate  the use of two different covariance matrices, hereafter known as  the standard Random-Walk (SRW) and Adaptive Random-Walk (ARW) proposal distribution. }
 
 In the SRW proposal distribution,  $\Sigma$ is fixed to be diagonal, so that $\Sigma=\mbox{diag}(\sigma^2_1,\ldots, \sigma^2_P)$; where $\sigma_j$ is the step size of the $j^{th}$ element of the parameter vector $\bs\theta$.  The step-size for parameter $\theta_j$  is chosen to be a
 combination of a fixed step size $\phi$ which is  common to all parameters,
 multiplied by the range of possible values for parameter $\theta_j$, so that

 \begin{equation}
 \sigma_j = (a_j - b_j) \times \phi
 \label{eq_stepsize}
 \end{equation}
 
 where, $a_j$ and $b_j$ represent the maximum and minimum limits of the prior for $\theta_j$   given in Table \ref{tab:truevalues}.

\textcolor{black}{ In the ARW proposal distribution, $\Sigma$ is adapted every $M$ intervals of within-replica sampling. $\Sigma$ allows for the dependency between elements of $\bs\theta$ and changes throughout the  within replica proposals  sampler \citep{haario2001adaptive}.  The elements of $\Sigma$ are  adapted to the posterior using the sample covariance of the current chain history:  $\Sigma = \mbox{cov}(\{\bs\theta^{[0]}, \ldots, \bs \theta^{[i-1]}\}) + \mbox{diag}(\lambda_1^2,\ldots,\lambda_P^2)$; where $\bs\theta^{[i]}$ is the $i^{th}$ iterate of $\bs\theta$ in the chain and $\lambda_j$ is the minimum allowed step sizes for each parameter $\theta_j$. }

 Given the current value  $\bs\theta^c$,  the proposed value for $\bs\theta$ denoted by $\bs\theta^p$, is accepted with probability  $\alpha$. This with uniform priors  and symmetric proposal distribution reduces to;

 \begin{equation}
 \alpha=\min\left(1,\frac{L(\mb D_T,\mb Z|\bs \theta^p)}{L(\mb D_T, \mb Z|\bs \theta^c)}
 \right)
 \label{eq_Like}
 \end{equation}

  \subsubsection{Between-Replica Sampling}
  The \textit{replica transition}  procedure  considers  the exchange of two neighboring  
replicas. Suppose there are $M$ replicas indexed by $m$, 
with corresponding stationary distributions  $p_m(\theta|\mb D_T)=p(\theta|\mb 
D_T)^{\beta_m}$; for, $m=1,\ldots,M$, with  $\beta_1=1$ and 
$\beta_M<\beta_{M-1}<\ldots,\beta_1$ and define the temperature ladder to be   $\bs\beta=(\beta_M,\ldots,\beta_1)$. The total parameter space consists of the chain $m$ and the parameters within those replicas $\theta_m$ and then  the pair $(m, \theta)$ are jointly
proposed and accepted/rejected according to the Metropolis-Hasting (MH) algorithm.
The stationary distribution of chain $m$ is
$p(\theta_m|\mb D_t,\mb Z)^{\beta_m}$.  Suppose chains $m$ and $m+1$ are at iteration $k$, with parameter values $\theta_{m}^{[k]}$  and $\theta_{m+1}^{[k]}$, which we denote  as $\theta_{m}^{c}$  and $\theta_{m+1}^{c}$; where, the superscript $c$ represents the current value. We propose a swap between these chains so that $\theta_{m}^{p}=\theta_{m+1}^{c}$  and $\theta_{m+1}^{p}=\theta_{m}^{c}$; where, the superscript $p$ represents the proposed  value.  We accept the  proposed values with probability
\begin{eqnarray}
\alpha & = & \min\left\{1,\left[\frac{p(\theta_{m}^{c}|\mb D_T,\mb Z)}{p(\theta_{m+1}^{c}|\mb D_T,\mb Z)}\right]^{(\beta_{m+1}-\beta_{m})}\right\}
\label{eq_swapacc}
\end{eqnarray} 
If accepted, $\theta_{m+1}^{[k+1]}=\theta_{m+1}^{p}$  and $\theta_{m}^{[k+1]}=\theta_{m}^{p}$; otherwise, the chain remains where it is, $\theta_{m+1}^{[k+1]}=\theta_{m+1}^{[k]}$  and $\theta_{m}^{[k+1]}=\theta_{m}^{[k]}$.
We note that each of the replicas are assigned with a temperature ladder that relaxes the likelihood  which affects the MH acceptance probability.    Essentially, the replicas with higher temperature values have more probability of within replica proposal acceptance which can help  in escaping from sub-optimal modes.  Given that the  temperature ladder is 
user-defined, it is dependent upon the nature of the posterior. We use a geometric temperature ladder with
\begin{equation}
\beta_m=1/\tau_m,\;\beta_M=1/M\;  \mbox{and therefore,}\; \beta_m=\beta_{m-1}M^{-1/(M-1)}
\label{temp_ladder}
\end{equation}
for $m=2,\ldots,M$ where $M$ is the number of replicas and $\tau$ is the maximum temperature which is user defined. 

\subsection{Multi-core   parallel tempering Bayeslands} 


 Bayeslands is comprised of the Badlands forward model embedded in a  PT-MCMC sampling scheme, both of which are computationally expensive,  and thus the combination of the two presents several  challenges. We note that in PT-MCMC, the replica with temperature level of 1 only becomes part of the posterior. The rest of the replicas are used to enhance \textcolor{black}{exploration} by escape from local minima. It is via the replica exchange  that the configuration  from the rest of the replicas become part of the posterior, provided that they move to replica with temperature of 1. It is  unfeasible to draw large number of samples given that the Badlands model is computationally expensive and hence the likelihood evaluation adds to the overall cost of sampling. Hence, we need need efficient sampling within limited time, given by number of samples drawn.


 
\begin{figure}[htb!]
  \begin{center}
    \includegraphics[width=150mm]{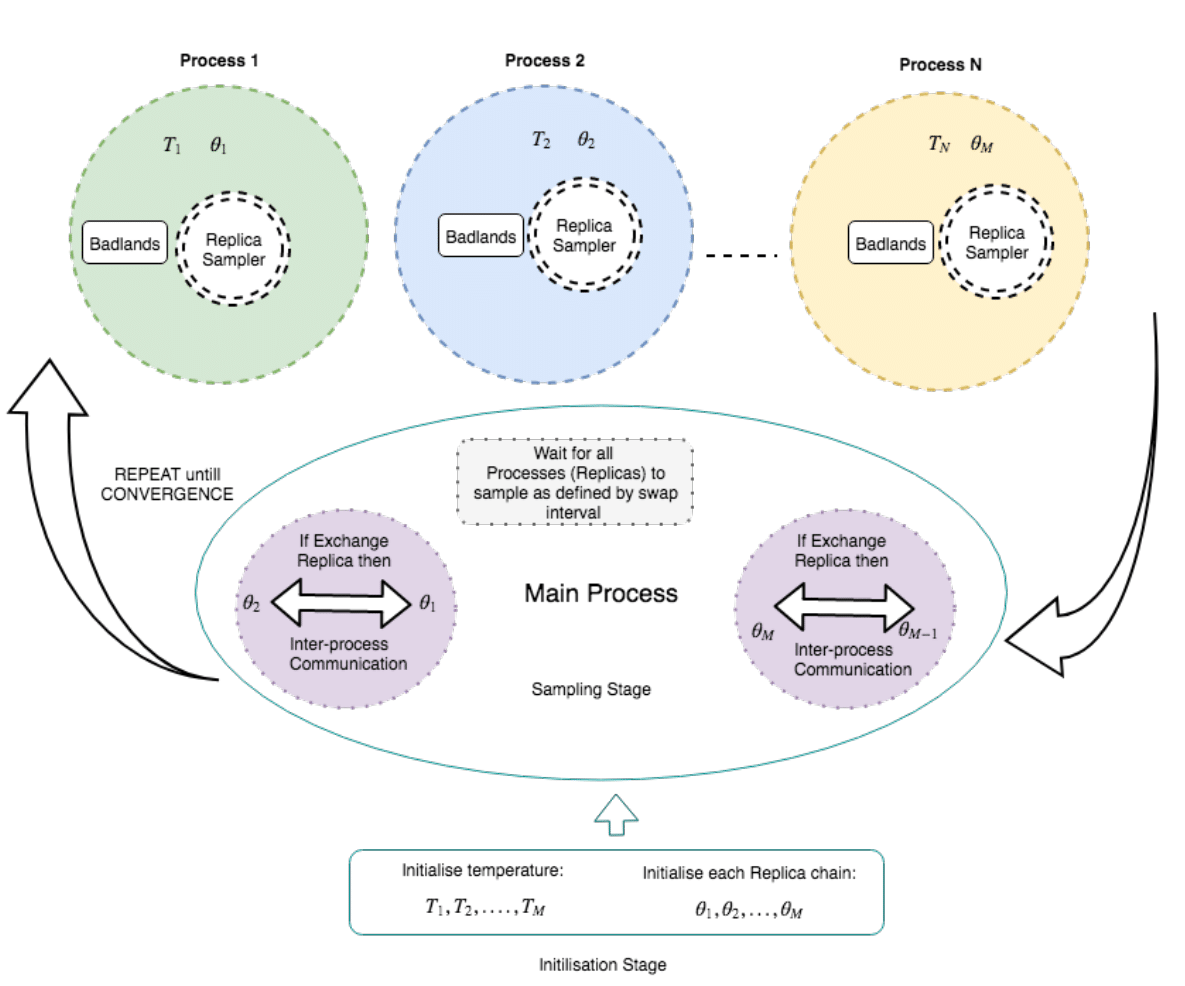} 
    \caption{An overview of  the different replicas that are executed on a 
multi-processing architecture for PT-Bayeslands. Note that the main process controls the given 
replicas and enables them to exchange the neighboring replicas.  }
 \label{fig:pt}
  \end{center}
\end{figure}

 To address this problem, we split the sampling scheme into two phases.  The first phase uses parallel tempering  with a temperature ladder as defined by Equation~\ref{temp_ladder}. The second phase sets the temperature ladder to be a vector of ones.  Essentially, the first phase is used to ensure that the starting values of the second phase cover most of the parameter space.  The second phase still  exchanges  between adjacent chains to avoid them getting stuck in local modes.   Figure \ref{fig:pt} gives an overview of  the different replicas that are 
executed on a multi-processing architecture.  The 
task of the main process in Figure \ref{fig:pt} is to manage the ensemble of replicas which runs on a separate processing core. Given  there are $M$ 
replicas  with each replica running on a 
separate core, there will be $M+1$ processes in total.  The Badlands model  is executed in the same  processing core as the within replica sampling, and inter-process communication is used to exchange neighboring 
replicas \citep{lamport1986interprocess}. To enhance computational efficiency, we minimize inter-process communication by only allowing replica exchanges at user defined fixed intervals. The main process waits for all replicas to complete sampling until the swap interval is reached   in order to 
attempt replica exchange.    Then the main process 
notifies the replicas to resume sampling with latest 
configurations in the chain for each replica.

Algorithm \ref{alg:mhptmcmc} outlines the sampling scheme.  It first initializes the number of replicas $M$, the maximum  number of iterations, $\mbox{Iter}_{\max}$, the swap interval $Swap_{int}$, and the temperature ladder $\bs\beta=(\beta_M,\ldots,\beta_1)$ before drawing the initial values of $\theta_m$ for $m=1,\ldots,M$ from the prior $p(\theta)$. Then, Algorithm \ref{alg:mhptmcmc} executes each of the replicas in parallel as in Figure \ref{fig:pt}. Each replica $\theta$ is updated when the respective proposal is accepted/rejected using the  MH acceptance criterion given by Step 1.4 and the new value of $\theta$ is added to the posterior distribution as shown in Step 1.3. This procedure is repeated until the replica swap interval is reached ($Swap_{int}$) which determines how often the algorithm pauses and checks if neighbouring replicas can be swapped using MH criterion (Step 2.3). The procedure is repeated until the termination condition is satisfied which is given by the maximum number of iterations. 
 
 It is 
generally expected that increasing the number of replicas in PT-Bayeslands will shorten the computation time; however, this is not necessarily true because additional time is taken  with  inter-process communication. The 
 effect of increasing the number of replicas on the  performance accuracy will be investigated.
\begin{algorithm}[H]
\small
\caption{Multi-core  Parallel Tempering MCMC }\label{alg:mhptmcmc}

\KwResult{Draw iterations from $p(\theta|\mb D_T,\mb Z)$}
i. Set maximum number of iterations  ($\mbox{Iter}_{\max}$), the swap interval ($\mbox{Swap}_{\mbox{int}}$), the number of replicas ($M$), the percentage of iterations for parallel tempering phase and the temperature ladder $\bs\beta=(\beta_M,\ldots,\beta_{1})$. \\
ii. Initialize replica $\theta_m=\theta_{m^{[0]}}^{[0]}$, \; for $m=1,\ldots,M$.\\
iii. \textcolor{black}{Set the current values;   $\theta_m^c=\theta_{m^{[0]}}^{[0]}$, and   $m^c=m^{[0]}$\;}\\
\While {$\mbox{Iter}_{\max}$}
{
 
 \For {$m=1,\ldots,M$   }{
 \For {$k=1,\ldots,$ $\mbox{Swap}_{\mbox{int}}$}{
 \noindent
 
*Parallel Tempering Phase \\
\If{tempering is false}{
$\bs\beta = 1_M$\;
}
*Parallel Tempering Phase \\

 Step 1: Replica sampling \\
 1.1 Propose $\theta^p$\\
1.2 Compute acceptance probability as in Equation~\ref{eq_Like}\; 

1.4 Acceptance criterion
Draw $u\sim U[0,1]$\;\\ 
\eIf{$u<\alpha$}{
$\theta^{[k]}=\theta^p$\;
}{
$\theta^{[k]}=\theta^c$\;
}
 
}
 Step 2: Replica Exchange \\
  2.1  Propose replica swap .\\
2.2 Compute acceptance probability as given by Equation~\ref{eq_swapacc}\; 

2.3 \textcolor{black}{Acceptance} criterion \\
Draw $u\sim U[0,1]$\;\\
 
\eIf{$u<\alpha$}{
$\theta_m^{[k]}=\theta^p_{m^p}$\;
}{
$\theta_m^{[k]}=\theta_{m^c}^c$\;
}

}
 
}
 \end{algorithm}

\subsection{Evaluation}
\label{S:2}

 The performance of PT-Bayelands is evaluated in three ways.  The first is by comparing the posterior distributions of the parameters with those used to generation the  data. The second is by comparing the computational time for the PT-Bayeslands versus single chain MCMC  (SC-Bayeslands) from earlier work \citep{CHANDRA2019}. The third  is by comparing the  predicted/simulated Badlands  landscape with the ground-truth data using  the root-mean squared error (RMSE). The RMSE  for the  elevation (elev) and sediment erosion/deposition (sed) is computed at each iteration of the sampling scheme and  are given by

\begin{eqnarray*}
 \mbox{RMSE}_{elev} & = &\sqrt{\frac{1}{n\times m}\sum_{i=1}^n\sum_{j=1}^n 
\left(g(\hat{\theta}_{T,i,j}) -g_{T,i,j}(\theta)\right)^2}\\
 \mbox{RMSE}_{sed} & = & \sqrt{\frac{1}{n_t \times 
v}\sum_{t=1}^{n_t}\sum_{j=1}^m \left(f(\hat{\theta}_{t,j}) 
-f(\theta_{t,j})\right)^2}
  \end{eqnarray*}

 \textcolor{black}{where, $\hat{\theta}$ is an estimated value of $\theta$, 
chosen according to 
\textcolor{black}{the proposal distribution} and $\theta$ is the true value  on which the ground truth 
topographies  and sediment thickness were based.  $f(.)$ and $g(.)$ represent 
the outputs of the  Badlands model, as defined earlier while $m$ and $n$ 
represent the size of the selected topography.  $v$ is the total number of 
selected points from   sediment erosion/deposition as shown in Figure  
\ref{fig:sediments} over the selected time frame, 
$n_t$. } Apart from the three  application problems (Cr, Mt, CM), we  also test the performance  of PT-Bayeslands to the different user settings, such as  exchange rate $\mbox{Swap}_{int}$, and the maximum number of iterations $\mbox{Iter}_{max}$. 

\section{Results}


\subsection {Estimated Topography  Elevation and Sediment  Erosion/Deposition}
\textcolor{black}{We show the estimates of the posterior mean elevation and the sedimentary erosion/depositions produced by PT-Bayeslands for all three landscape problems and compare these with the ground-truth.  For all three problems, the number of iterations was fixed at 10,000  with 10 replicas, and a SRW was used as a proposal distribution. PT-Bayeslands produces posterior means of the elevation and sedimentary deposits each year but due to limited space, we show these figures for only two selected years; one half way through the evolution and the other at the end of the evolution.}

\begin{figure}[htb!]
  \begin{center}
    \includegraphics[width=140mm]{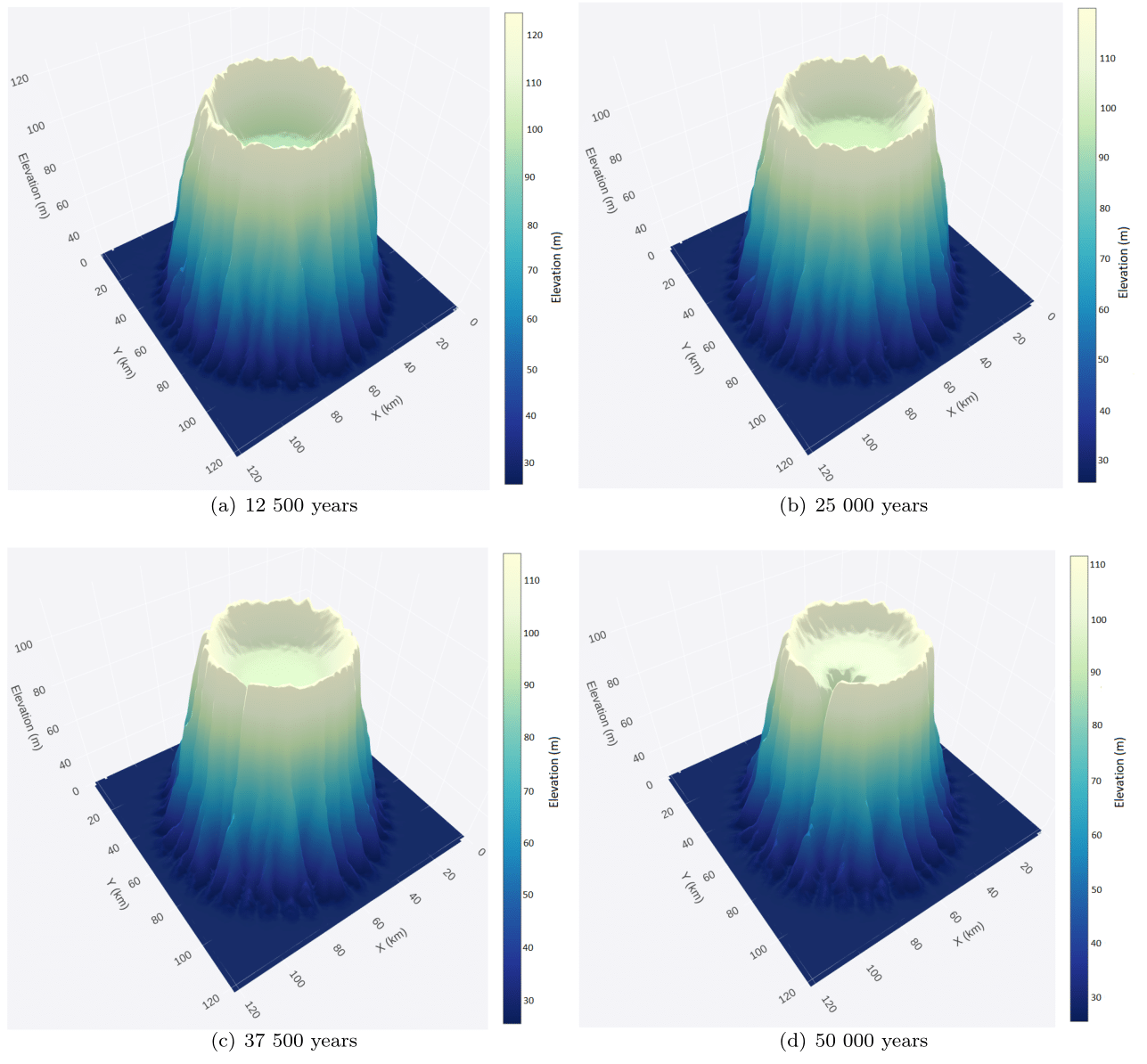} 
 \caption{Badlands landscape evolution for 4 different time-scales for the Synthetic-Crater showing the elevation topography, for 10 000 iterates with 10 replicas, using  simple random-walk  as a proposal distribution from   results shown in  Table  \ref{tab:all_res}.}
 \label{fig:elev_cr} 
  \end{center}
\end{figure}

\begin{figure}[htb!]
  \begin{center}
    \includegraphics[width=140mm]{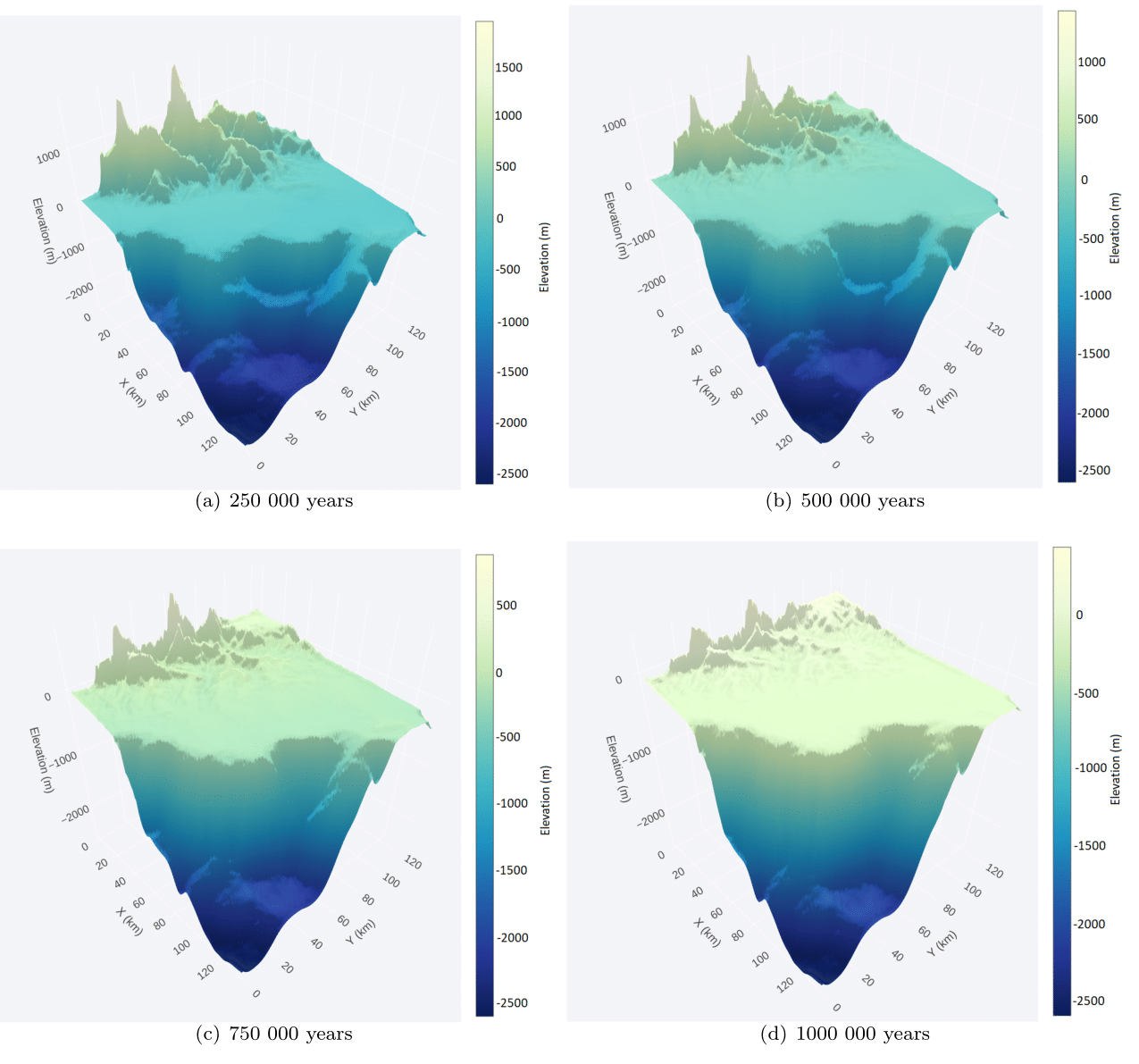} 
 \caption{Badlands landscape evolution for  4 different time-scales for the Continental-Margin showing the elevation topography  for 10000 iterates with 10 replicas, using simple random-walk as a proposal distribution from   results shown in  Table  \ref{tab:all_res}. }
 \label{fig:elev_cm}
  \end{center}
\end{figure}

\begin{figure}[htb!]
  \begin{center}
    \includegraphics[width=140mm]{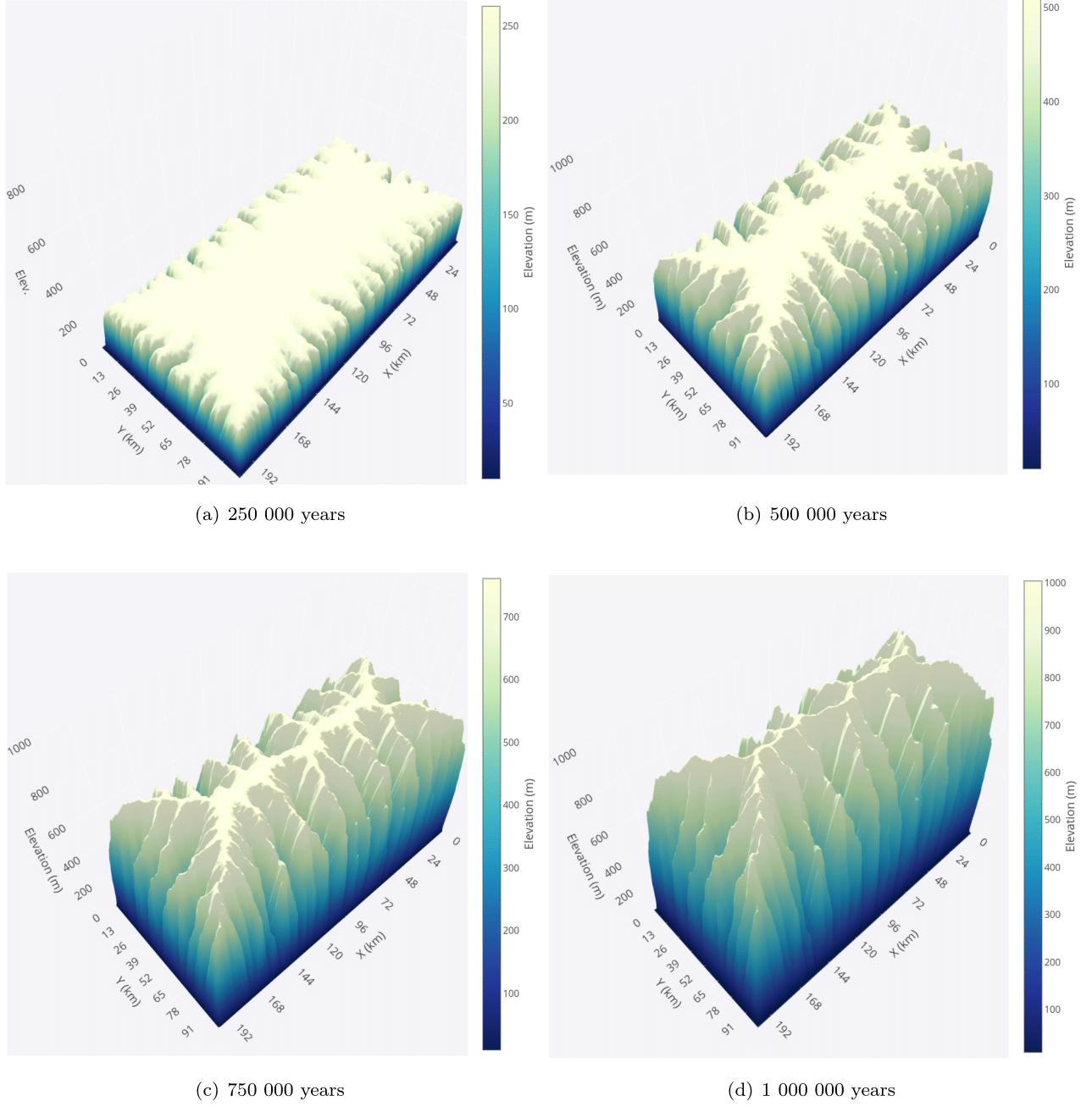} 
  \caption{  Badlands landscape evolution for  4 different time-scales for the Synthetic-Mountain showing the elevation topography  for 10 000 iterates with 10 replicas, using simple random-walk as a proposal distribution from   results shown in  Table  \ref{tab:all_res}.}
 \label{fig:elev_mt}
  \end{center}
\end{figure}

\textcolor{black}{Figure~\ref{fig:elev_cr} shows the elevation of the Cr problem after 25,000 and 50,000 years.  Figures~\ref{fig:elev_cm}~and~\ref{fig:elev_mt}, are analogous plots for the CM and Mt problems for 500,000 years and   1,000,000 years, respectively . These plots show how well PT-Bayeslands estimates the final elevation ground truth.}

\begin{figure}[htb!]
  \begin{center}
    \includegraphics[width=140mm]{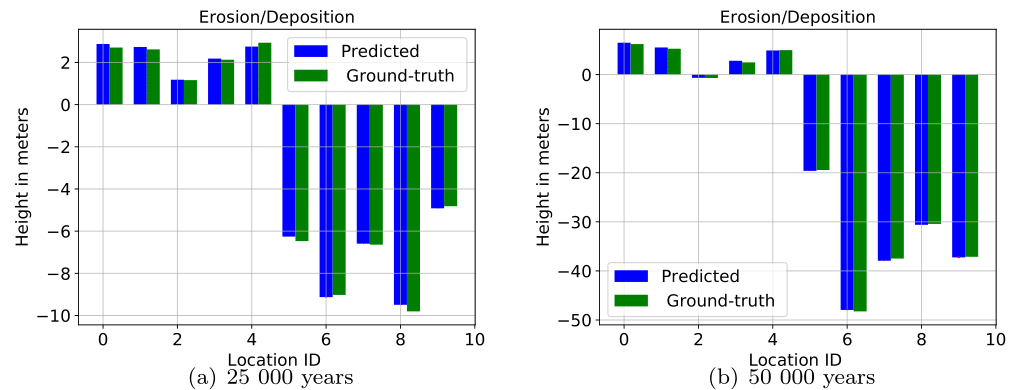} 
    \caption{  Badlands landscape evolution for  2 selected  time-scales corresponding to Figure \ref{fig:elev_cr} for the Synthetic-Crater showing the sedimentary deposition for 10 selected points. }
 \label{fig:sediments_cr}
  \end{center}
\end{figure}

\begin{figure}[htb!]
  \begin{center}
    \includegraphics[width=140mm]{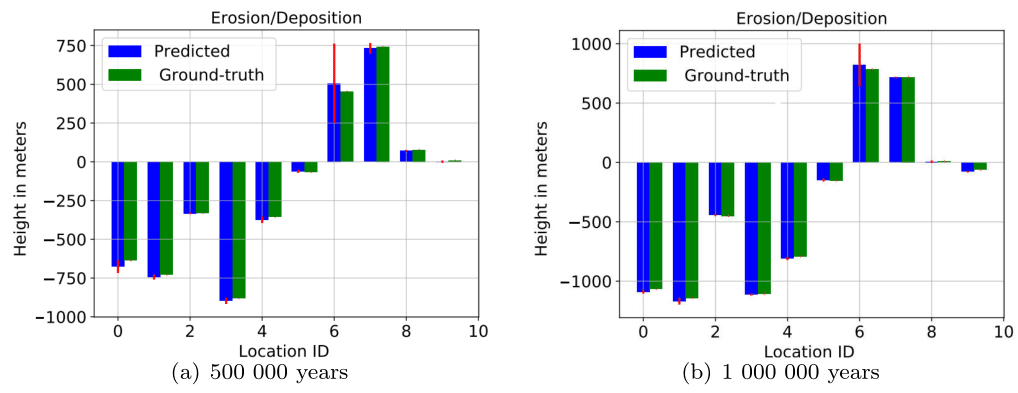} 
    \caption{  Badlands landscape evolution for  2   selected  time-scales corresponding to Figure \ref{fig:elev_cm} for the Continental-Margin showing the sedimentary deposition for 10 selected points. \textcolor{black}{The red lines show  the uncertainty in prediction given by 95 \% credible interval.} }
 \label{fig:sediments_cm}
  \end{center}
\end{figure}

\begin{figure}[htb!]
  \begin{center}
    \includegraphics[width=140mm]{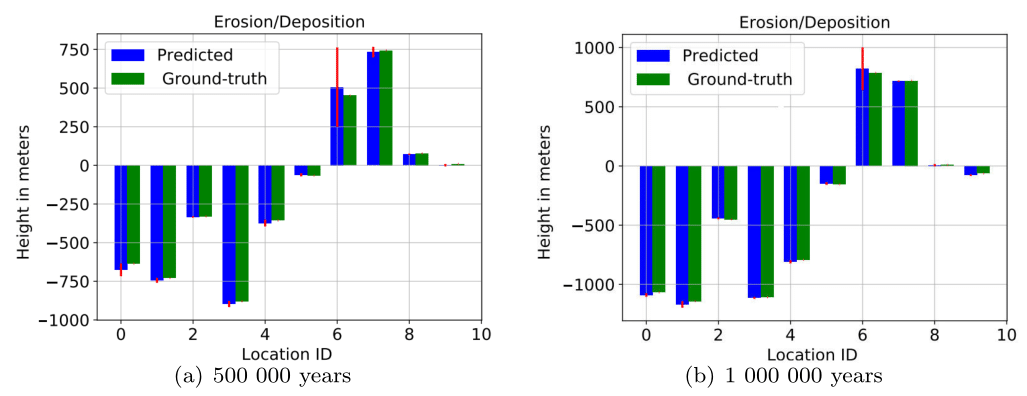}
    \caption{  Badlands landscape evolution for   2   selected  time-scales corresponding to Figure \ref{fig:elev_mt} for the Synthetic-Mountain showing the sedimentary deposition for 10 selected points. \textcolor{black}{The red lines show  the uncertainty in prediction given by 95 \% credible interval.} }
 \label{fig:sediments_mt}
  \end{center}
\end{figure}

\textcolor{black}{Figures~\ref{fig:sediments_cr}, \ref{fig:sediments_cm} and \ref{fig:sediments_mt} show the sedimentary erosion/deposition at the same two selected time slices.  These sedimentary erosion/deposition data were taken at 10 selected points across the given landscape (Figure \ref{fig:sediments}).  The posterior mean is shown in black while the ground truth is shown in green. These figures show how well PT-Bayeslands approximates the sedimentary erosion/disposition when compared to the  ground truth at   selected times.}

\subsection{Multi-modality  }

\begin{figure}[htb!]
  \begin{center}
    \includegraphics[width=140mm]{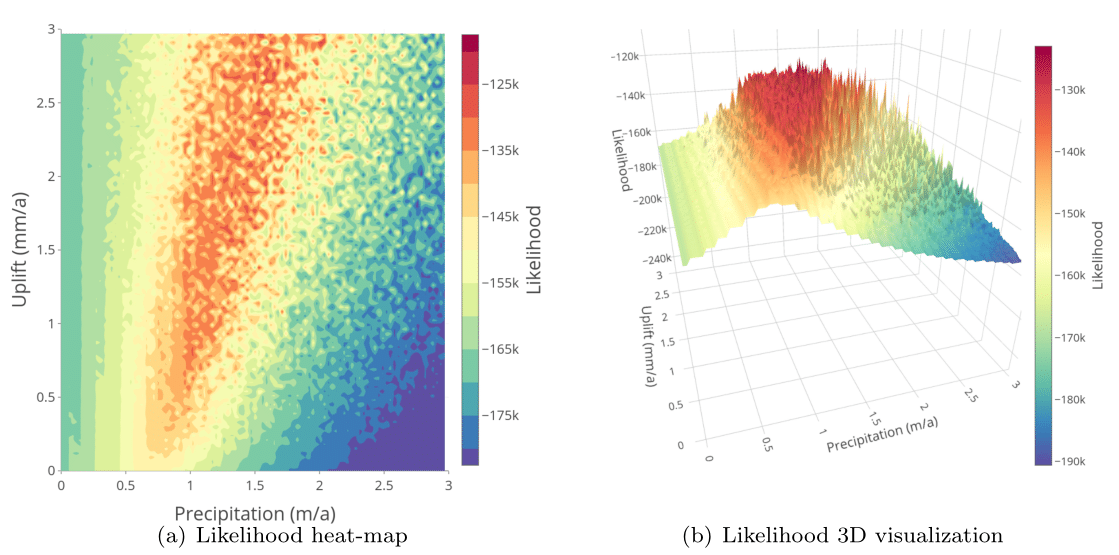}
    
    \caption{Panels~(a)~and~(b) are a heat map and a surface plot of the log posterior  surface of the Mountain problem  as function of   precipitation, $\rho, $ and erodibility, $k$.}
 \label{fig:lks_mountain}
  \end{center}
\end{figure}

 Geophysical inversion problems often exhibit parameter posterior distributions which are multi-modal.  Figure ~\ref{fig:lks_mountain} shows the log posterior for  precipitation and  uplift while fixing all other parameters in the Mt problem.  We again note that our previous work  \citep{CHANDRA2019} showed that  different combinations  of precipitation  and erodibility  gave rise to visually indistinguishable topography for both the Cr and CM problem.   
 
 Figure~\ref{fig:lks_mountain} shows that the posterior has a ridge-like structure which is  similar to Figure~\ref{fig:lks_craterx}; however,  the surface exhibits so many discontinuities that resemble CM posterior surfece in Figure \ref{fig:lks_cm}. We found  that SC-Bayeslands had difficulty  to uncover or efficiently sample the structure.  In this case, even PT-Bayeslands will face  difficulties and carefully crafted proposal distribution and  temperature ladder will be required to efficiently explore the space.

 \subsection{Performance comparison  for simpler problems } 
 \label{sec:SC_vs_PT}
 
 To further investigate the issue of multi-modality,  we compared PT-Bayeslands with SC-Bayeslands for the respective problems  in different settings.  In the first setting, we allowed only one parameter at a time to vary and fixed the other parameters to their true value.  The parameters that were allowed to vary were precipitation ($\rho$) and erodibility $k$ for the Cr and CM problem while rest were fixed to true values shown in Table \ref{tab:truevalues}. In the second setting, $\rho$ and $k$ were allowed to vary simultaneously while fixing the remaining parameters.  
 
\begin{figure}[htb!]
  \begin{center}
    \includegraphics[width=140mm]{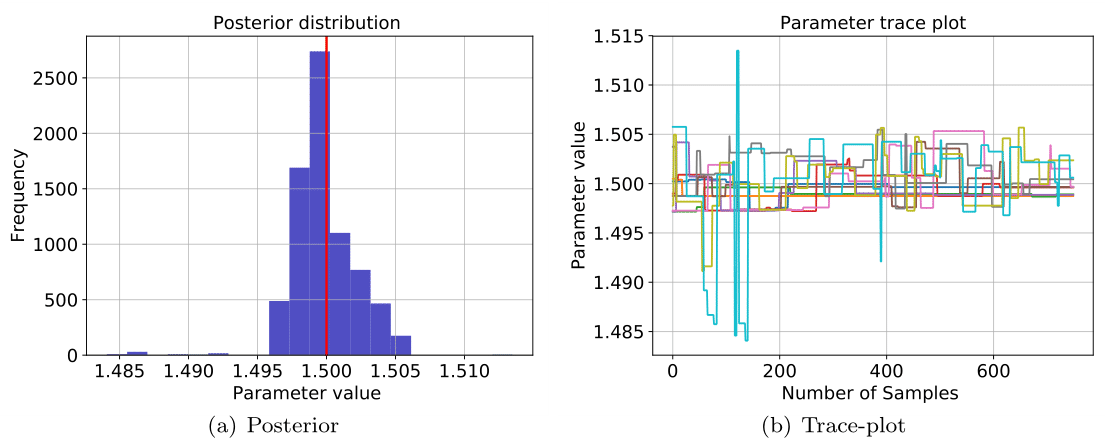}
    
    \caption{ PT-Bayeslands for Cr problem with one parameter (precitipation) for  10, 000 iterations. The panels show  estimates of the posterior distribution and trace-plot of the precipitation ($\rho$). \textcolor{black}{The trace-plots show the accepted or current values of the chain for the given samples}. The true value is given by red vertical line (1.5 m/yr) in Panel (a).   }
 \label{fig:oneparam_crater}
  \end{center}
\end{figure}

 Figure~\ref{fig:oneparam_crater} shows histogram estimates of the posterior distribution for $\rho$,  and trace plots of the iterates of $\rho$ from PT-Bayeslands for the Cr problem, while fixing all the other parameters.  The true value of $\rho$ is 1.5 m/a  which is given by the vertical red line that appears in the centre of the distribution. The trace plot shows that the replicas mix well. Similar plots for the other parameters and on the CM problem were also obtained.
 
\begin{figure}[htb!]
  \begin{center}
    \includegraphics[width=140mm]{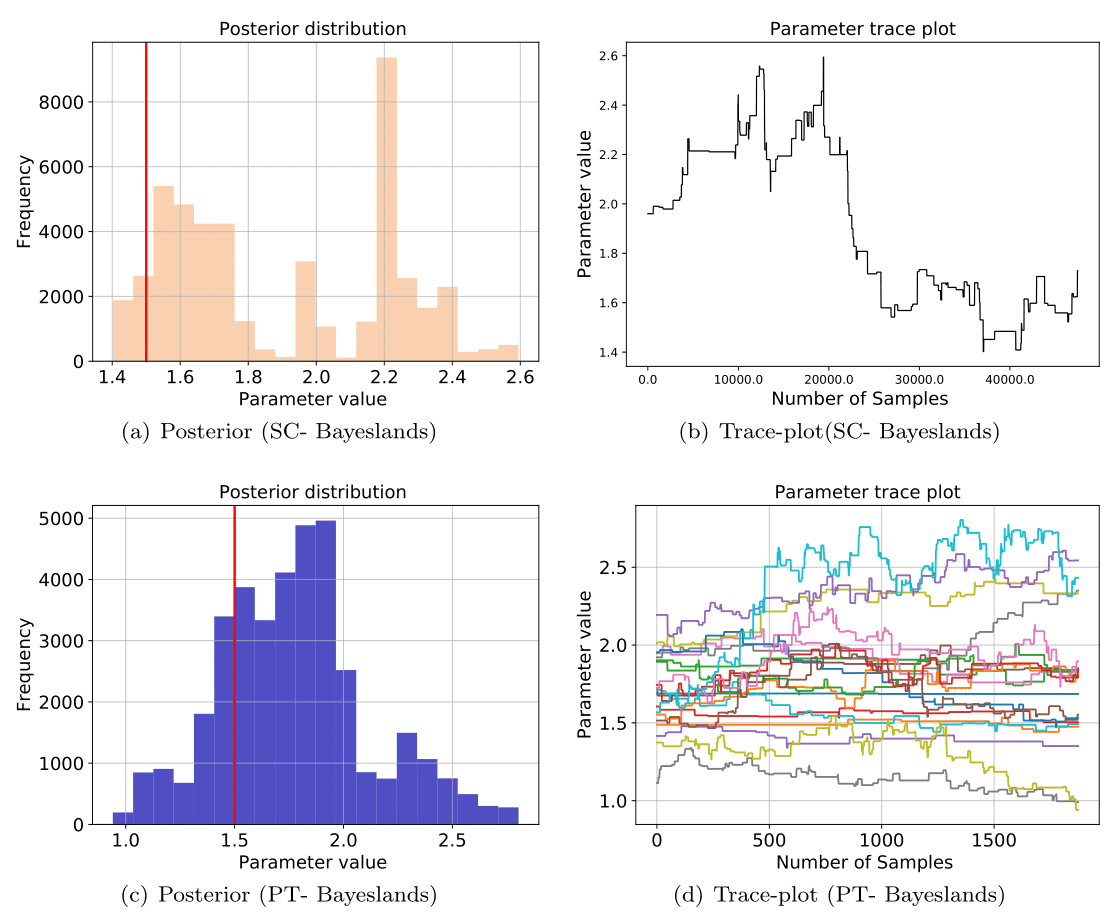}
    
    \caption{Comparison of PT-Bayeslands with SC-Bayeslands for the Cr problem for 50,000 iterations. The panels show  estimates of the posterior distribution and trace-plot of the precipitation ($\rho$) for the SC-Bayeslands and PT-Bayeslands respectively. \textcolor{black}{The trace-plots show the accepted or current values of the chain for the  given samples}. Note that the results for SC-Bayeslands are taken from  \citep{CHANDRA2019} (with 5 \% burnin) while PT-Bayeslands features 25 \%burn-in.  }
 \label{fig:crater_posSC-PT}
  \end{center}
\end{figure}

 Figure~\ref{fig:crater_posSC-PT} displays results for the Cr problem in the second setting. Figure~\ref{fig:crater_posSC-PT}, Panels~(a)~and~(c,) show histogram estimates of the posterior distribution of $\rho$, when both $\rho$ and $k$ are allowed to vary.  Panel~(a) displays the estimated histogram using  SC-Bayeslands, while panel~(c) displays the estimated histogram using PT-Bayeslands. Panels~(b)~and~(d) are the corresponding trace plots. These plots show how PT-Bayeslands vastly improves the exploration of the parameter space. The sampler in SC-Bayeslands explores one mode for the first 20,000 iterations and then appears to shift to another mode for the remaining iterations.  In contrast, PT-Bayeslands explores the values of $\rho$ corresponding to the entire ridge in Figure~\ref{fig:lks_craterx}.
 
\begin{figure}[htb!]
  \begin{center}
    \includegraphics[width=140mm]{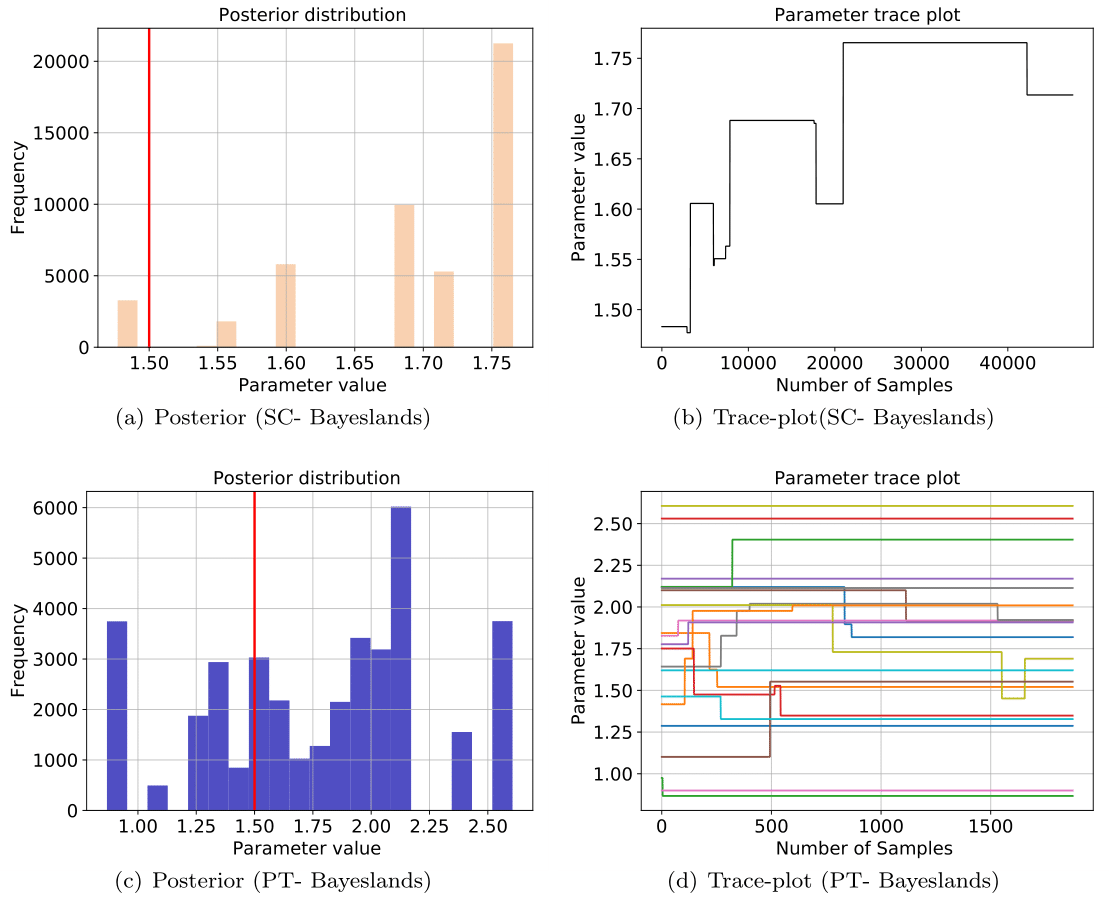}

    \caption{Comparison of PT-Bayeslands with SC-Bayeslands for the CM problem for 50,000 iterations. The panels show  estimates of the posterior distribution  and trace-plot of the precipitation ($\rho$) for the SC-Bayeslands and PT-Bayeslands respectively. \textcolor{black}{The trace-plots show the accepted or current values of the chain for the given samples}.   Note that the results for SC-Bayeslands is taken from  \citep{CHANDRA2019} (with 5 \% burnin) while PT-Bayeslands features 25 \%burn-in.  }
 \label{fig:cm_posSC-PT}
  \end{center}
\end{figure}

 Figure~\ref{fig:cm_posSC-PT} displays  results for the CM problem which  shows a similar trend. Although PT-Bayeslands performs better than SC-Bayeslands  (Figure~\ref{fig:cm_posSC-PT}), the mixing between chains is still poor. This is due to the large number of local modes in the log-posterior surface shown in Figure  \ref{fig:lks_cm}, which earlier highlighted the challenges of sampling.  Figure  \ref{fig:lks_cm} shows that the number of modes is much greater than the number of chains, and a possible solution to this is to add more chains, but then the sampler is not computationally feasible.

\begin{table*}[htbp!]
\centering
\smaller
 \begin{tabular}{ l c c c c c c } 
 \hline 
 Problem &Method   & Parameters & Time &  $RMSE_{elev}$ & $RMSE_{sed}$ \\        
&  &allowed to vary & (minutes) &  pos. mean  &   pos. mean  \\  
   \hline  
 Synthetic-Crater & SC-Bayeslands  &   $\rho$  &	 355.92	& 1.040& 0.179  \\ 
 & SC-Bayeslands  &  $k$  &	356.18  & 1.05  & 0.180 \\ 
 & SC-Bayeslands  & $\rho$, $k$  &  355.39 & 1.05 & 0.170 \\
 
& PT-Bayeslands &   $\rho$ & 44.05    & 1.050 & 0.180 \\
& PT-Bayeslands &  $k$ &    42.31   & 1.050 & 0.180 \\
& PT-Bayeslands & $\rho$, $k$  &   39.91  & 1.050 & 0.180 \\
 
\hline
Continental-Margin& SC-Bayeslands  & $\rho$ &	418.0	& 52.4  	& 34.1  \\ 
& SC-Bayeslands  & $k$ &	423.70 & 51.50	& 43.30  \\ 
&SC-Bayeslands  & $\rho$, $k$  & 415.80 & 44.00   & 48.90 \\
 
& PT-Bayeslands & $\rho$ &   59.40   & 58.3,   & 48.10 \\
& PT-Bayeslands & $k$ & 64.70     & 59.6  & 46.90 \\
&PT-Bayeslands & $\rho$, $k$  & 60.60     & 60.3 & 50.60 \\

 \hline 
 \end{tabular}
 
\caption{Comparison of SC-Bayeslands and PT-Bayeslands for the CM and Cr problems. The parameter column presents  the parameters that are estimated while keeping other  model parameters  fixed to their true values. The time taken for  SC-Bayeslands and PT-Bayeslands is reported along with the accuracy given as posterior mean (pos. mean) of $\mbox{RMSE}_{\mbox{elev}}$ and  $\mbox{RMSE}_{\mbox{sed}}$.  
}
 \label{tab:initial} 
\end{table*}

  Table \ref{tab:initial} shows the results from the experiments for the different combination of  parameters  for the two methods. Note that the total prediction accuracy is given for model elevation topography and sedimentary thickness in terms of accuracy given as posterior mean of $\mbox{RMSE}_{\mbox{elev}}$ and  $\mbox{RMSE}_{\mbox{sed}}$.    The sediment RMSE  considers the ground-truth and final sediment distribution averaged over selected points in the model domain (see Figure \ref{fig:sediments}). The results show that  both methods achieve  equally consistent prediction accuracy. However, PT-Bayeslands  significantly reduces the computational  time taken while also having better exploration features as shown in Figure \ref{fig:crater_posSC-PT}. 


\subsection{Effect of simulation setting}
\label{sec_SimSet}


In this section, the effect of the number of replicas, the number of iterations, and the swap rate are evaluated using the CM problem given all six free parameters as shown in Table \ref{tab:priors}. This problem was selected for these experiments  because it  has a difficult log-posterior surface with a single optimal mode and many sub-optimal modes.    Table \ref{tab:eval_cores} 
provides metrics for each of the settings, where each setting has a total number of 10,000 iterations.  For example, the first setting has 10000 iterations spread across 4 chains, therefore a total of 2500 for each while, while last setting  has 48 chains and therefore only 208 iterations in each chain.  We observe that increasing the number of 
replicas reduces the overall computation time. However, the number of replicas 
does not appear to have much effect on the elevation and sediment erosion/deposition prediction accuracy as measured by the $\mbox{RMSE}_{\mbox{elev}}$ and $\mbox{RMSE}_{\mbox{sed}}$. 

We then evaluate the effect that an increase in the number of iterations has on the prediction accuracy of PT-Bayeslands, for a given number of replicas.  \textcolor{black}{Table \ref{tab:res_etoposhort}  presents summary of the 
results  for the CM topography with  
10 replicas. Obviously, increasing the number of iterations   increases the 
 overall time taken in a linear fashion. The relationship between the number of iterations and the prediction accuracy is non-linear given  diminishing improvements to prediction accuracy as the number of iterations is increased.  The optimal setting depends upon the trade-off between computational speed and prediction accuracy. We find that   the interval 20,000 to 50,000 iterations is reasonable.}  
 
 \textcolor{black}{Table \ref{tab:swapratio} contains the results of an experiment to investigate the effect of the replica exchange rate  (Swap-Rate) on  the prediction accuracy and the percentage of swaps accepted  for the CM problem. Here, we used 10,000 iterations with 10 replicas. The results show that the Swap-Rate has a negligible effect on the  prediction accuracy. }

\begin{table}[htbp!]
\centering
\smaller
 \begin{tabular}{ c c c c c} 
 \hline 
  Replicas & Time & $\mbox{RMSE}_{elev}$& $\mbox{RMSE}_{sed}$\\  
 (cores) & (minutes)  & (pos. mean) &  (pos. mean)\\  
 \hline 
4	 &	153.2	 &	 58.6  	 &	 43.3 	 \\
8	 &	80.5	 &	 56.9 	 &	 48.7 	 \\
10	 &	63.4	 &	 60.4  	 &	 49.1 	 \\
12	 &	60.0	 &	 56.7  	 &	 47.9 	 \\
16	 &	43.4	 & 67.2   &	 52.3 	 \\
20	 &	35.3	 &	 67.6  	 &	52.8	 \\
24	 &	30.2	 &	 68.6  	 &	51.2	 \\
30	 &	26.1	 &	 69.0  	 &	51.5	 \\
36	 &	22.7	 &	 71.7  	 &	53.7	 \\
42	 &	20.1	 &	 71.1   &	53.0	 \\
48	 &	19.1	 &	 75.7  	 &	52.9  \\
 \hline
 \end{tabular}
 
\caption{Effect of  increasing the number replicas for the CM topography with 10\,000 total iterations. The  prediction accuracy is given as posterior mean (pos.mean)   of $\mbox{RMSE}_{\mbox{elev}}$ and  $\mbox{RMSE}_{\mbox{sed}}$ }
 \label{tab:eval_cores} 
\end{table}

\begin{table}[htbp!]
\centering
\smaller
 \begin{tabular}{ c c c c c} 
 \hline 

  iterations  & Time  & $\mbox{RMSE}_{elev} $& $\mbox{RMSE}_{sed}$\\  
  
  &  (minutes)  & (pos. mean ) &  (pos. mean)\\
 \hline   
1000	 &	6.5	 &	 80.6  	 &	 64.1  	 \\
5000	 &	29.4	 &	 67.5  	 &	 49.4  	 \\
10000	 &	54.1	 &	 61.5  	 &	 52.8  	 \\
20000	 &	130.5	 &	 57.3  	 &	 51.2   \\
50000	 &	279.3	 &	 49.0  &	 45.7   \\
100000	 &	518.3	 &	 48.6  	 &	 43.2  	 \\
 \hline
 \end{tabular}
 
\caption{Effect of the number of iterations for the CM  problem with 10 replicas. The  prediction accuracy is given as posterior mean (pos.mean)   of elevation $\mbox{RMSE}_{\mbox{elev}}$ and  sediment $\mbox{RMSE}_{\mbox{sed}}$ . }
 \label{tab:res_etoposhort} 
\end{table}

\begin{table}[htbp!]
\centering
\smaller
 \begin{tabular}{ c c c c c c c} 
 \hline 

  Swap-Rate  &Time  & $\mbox{RMSE}_{elev}$ &$\mbox{RMSE}_{sed} $& Swap \%  \\  
  
  & (minutes) &     (pos. mean) &  (pos. mean) &    \\
 \hline 

0.01  	&	88.4 &	 69.9 	&	51.2 	&	22.2 	\\
0.02 	&	94.6 &	 72.8 	&	58.6 	&	23.0 	\\
0.03	&	100.2 &	 76.7  	&	58.2 	&	21.8 	\\
0.04	&	91.5 &	 66.1  	&	46.5 	&	22.4 	\\
0.05		&	90.1 &	 63.9  	&	45.2 	&	22.9	 	\\
0.06	&	94.5 &	 74.5 	&	51.9 	&	23.4	 	\\
0.07	&	90.1 &	 69.7  	&	50.7 	&	22.0 	\\
0.08	&	92.5 &	 71.1  	&	51.7 	&	22.2	 	\\
0.09	&	88.5 &	 67.5  	&	44.3 	&	25.7	 	\\
0.10	&	95.2 &	 69.8  	&	58.2 	&	22.8 	\\

 \hline
 \end{tabular}
 
\caption{Effect of the swap-rate  for the CM problem with 10 replicas (10, 000 iterations in total). Note that the posterior mean (pos. mean) for accuracy in terms of $\mbox{RMSE}_{elev}$ and $\mbox{RMSE}_{sed}$ are  given in meters. The percentage swap (Swap \%) gives an indication of number of proposals that were successfully exchanged. The  prediction accuracy is given as posterior mean (pos. mean)   of elevation $\mbox{RMSE}_{elev}$ and sediment $\mbox{RMSE}_{sed}$.  }
 \label{tab:swapratio} 
\end{table}

\subsection{Effect of the Proposal Distribution}
 

In this section, we allow the number of iterations and the number of replicas to jointly vary, as well as allowing the proposal distribution (SRW and ARW).  The effect of these different settings on the prediction accuracy  is evaluated for all three landscape problems shown in Table \ref{tab:all_res}. A major observation in Table~\ref{tab:all_res} is that the proposal distribution  does not effect the prediction accuracy. 


\begin{table*}[htbp!]
\centering
\smaller
 \begin{tabular}{ l l l l l l l  l  l l} 
 \hline 

 Topography & Parameters & Proposals & Replica  & $T_{max}$s & $\mbox{RMSE}_{elev}$ & $\mbox{RMSE}_{sed}$ & Swap\%  \\  
  &  & & & &   (mean, std) &   (mean, std) & &\\  
 \hline  
 Cr&       4   &   SRW  &  10 & 10 000	 &   (1.05, 0.01)  	&  (0.18, 0.01) & 22.15   \\ 
 
   &   &	  & 20  & 50 000   &     (1.05,0.01) & (0.18, 0.01) &13.09    \\  
 
 &    & \textcolor{black}{ARW} 	 & 10 & 10 000 &      (1.05, 0.01) & (0.18,  0.01) 	& 33.66  \\
 
 &    &  	 & 20 & 50 000 &      (1.05,   0.01) & (0.18,  0.01)  	& 26.33 \\
 \hline

Mt&   5  &SRW & 10  &  10 000 &  (6.30, 3.11) & (0.87, 1.18)  & 22.31 \\

    &  &	& 20  &50 000      	& (5.48, 1.36) & (0.54, 0.51)  & 13.26   \\  

  &    &  \textcolor{black}{ARW } 	 & 10 & 10 000 &	    (6.69, 2.55)  & (0.69, 0.60)  	& 22.30  \\

  &    & 	 & 20 & 50 000 &	    (5.44,  1.72) &  (0.48,  0.37)  	& 13.26   \\ 
 
 \hline
 
 CM &  6  &SRW	 & 10  & 10 000 	 & (61.68, 17.31) &    (45.74, 25.31)	& 23.62  \\ 
 
  & & & 20  & 	50 000     	& (61.86, 8.64) & (48.38, 17.97) & 14.32   \\
 
 &    & \textcolor{black}{ARW }	 & 10  & 50 000	 & (70.56, 8.38) & (50.80, 12.50)	& 31.87  \\ 
  &    &   & 10  & 50 000	 & (63.02, 11.48) & (52.53, 20.81)	& 27.22  \\

 \hline
 \end{tabular}
 
\caption{Typical results in PT-Bayeslands using   simple random-walk (SWR) and adaptive random-walk (ARW proposal distributions for the three problems (Cr, Mt and CM). $T_{max}$ denotes the maximum number of iterations and the mean and standard deviation (std) of topography elevation ($\mbox{RMSE}_{elev}$) and sediment erosion/deposition ($\mbox{RMSE}_{sed}$) prediction accuracy is shown.  }
 \label{tab:all_res} 
\end{table*}


\begin{table*}[htbp!]
\centering
\smaller
 \begin{tabular}{ l l l l l l l l l l  } 
 \hline  

 Topography & Proposal &   $\rho$    & $k$  & $\gamma$  & $\lambda$ & $\beta$  & $\phi$ & $u$   \\  
  
 \hline  
 Cr&            SRW  &   4.71 &  5.19  & 2.63 & 11.49  & - &- &- \\  
 &     \textcolor{black}{ARW} 	 &  4.39  & 3.79 & 1.25&  1.13 & - &- &-    \\
  
 \hline

Mt&     SRW &   5.99 &  8.57 & 10.29  &13.62 & 12.55 &7.27 &- \\ 
  &      \textcolor{black}{ARW } 	 &  1.78 & 1.77 & 1.17&  1.33&  1.12& 1.24  &   - \\

 \hline
 
 CM &    SRW	 &   4.48  & 3.11  &2.63  &6.14  & - & -  & 2.81  \\  
 &      \textcolor{black}{ARW }	 &  1.73 &  1.89 & 1.16 & 1.34 & - & - & 1.21    \\

 \hline
 \end{tabular}
 
\caption{ Convergence diagnosis showing the  potential scale reduction factor  (PSRF) score for the simple random-walk (SWR) and adaptive random-walk (ARW proposal distributions for the three problems (Cr, Mt and CM) are shown. The respective  parameters include;  precipitation ($\rho$), erodibility ($k $), m-value ($\gamma$), n-value ($\lambda$), marine ($\beta$), surface ($\phi$) and uplift ($ u$).  }
 \label{tab:convergence} 
\end{table*}

\textcolor{black}{Table \ref{tab:convergence}  shows convergence results for the two different proposal distributions as measured by the Gelman-Rubin diagnostic \citep{gelman1992inference}. The Gelman-Rubin diagnostic evaluates MCMC  convergence by analyzing the behaviour of  multiple Markov chains. Given multiple chains from different experimental runs, assessment is done  by comparing the estimated between-chains and within-chain variances for each  parameter, where large differences between the variances indicate non-convergence.  We provide convergence diagnosis for the two different proposal distributions (SRW and ARW) by running multiple chains using  different initial values for 10,000 iterations for each topography problem. We  calculate the potential scale reduction factor (PSRF) which  gives the ratio of the current variance in the posterior variance for each parameter compared to that being sampled.   The values for the PSRF near 1 indicates convergence.     Table \ref{tab:convergence}  shows that the AWR proposals in general provide much better convergence  when compared to SRW proposals. The only case where ARW is not near 1 is for two parameters ($\rho$, $k $).   }

\begin{figure}[htb!]
  \begin{center}
    \includegraphics[width=140mm]{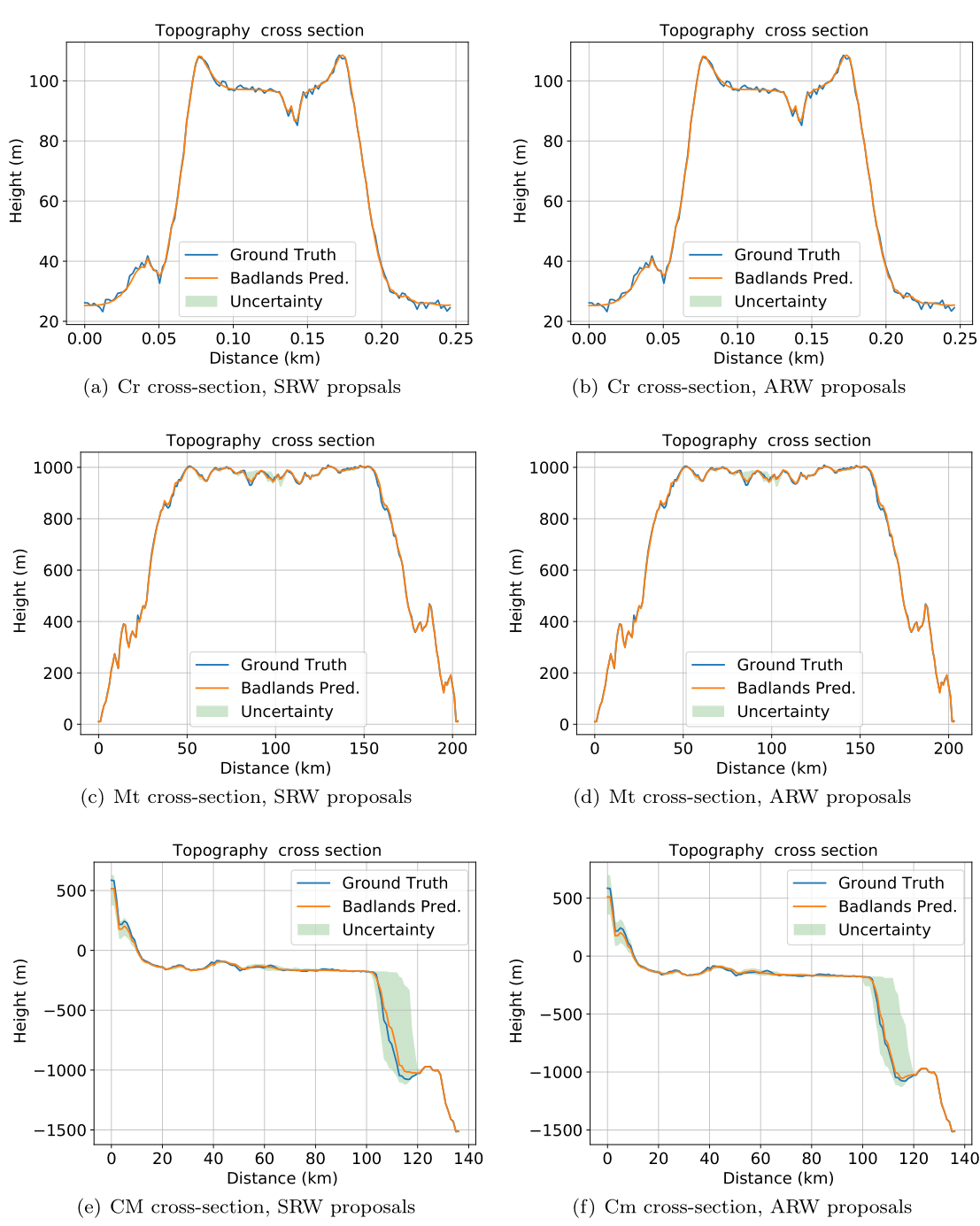}
    
    \caption{Cross-section   comparing the ground-truth evolved topography of the Badlands model with the PT-Bayeslands predictions for the respective problems  using SRW and ARW proposal. \textcolor{black}{Note that the x-axis for the subplots are different due to the dimensions of the respective problems}.}
 \label{fig:cross_section}
  \end{center}
\end{figure}

Figure~\ref{fig:cross_section} show the posterior mean and associated uncertainty of a cross section through the $Y$ axis, with the $X$ axis held constant at its middle value for both the SRW proposal and the ARW proposal for all the three problems.    The figure shows that the estimated cross sectional elevation is  visually identical for both types of proposals, as expected theoretically.  We note that the uncertainty for the Mt problem is slightly less for the ARW than the SRW, suggesting that the SRW may not have converged as quickly.  

\begin{figure}[htb!]
  \begin{center}
    \includegraphics[width=110mm]{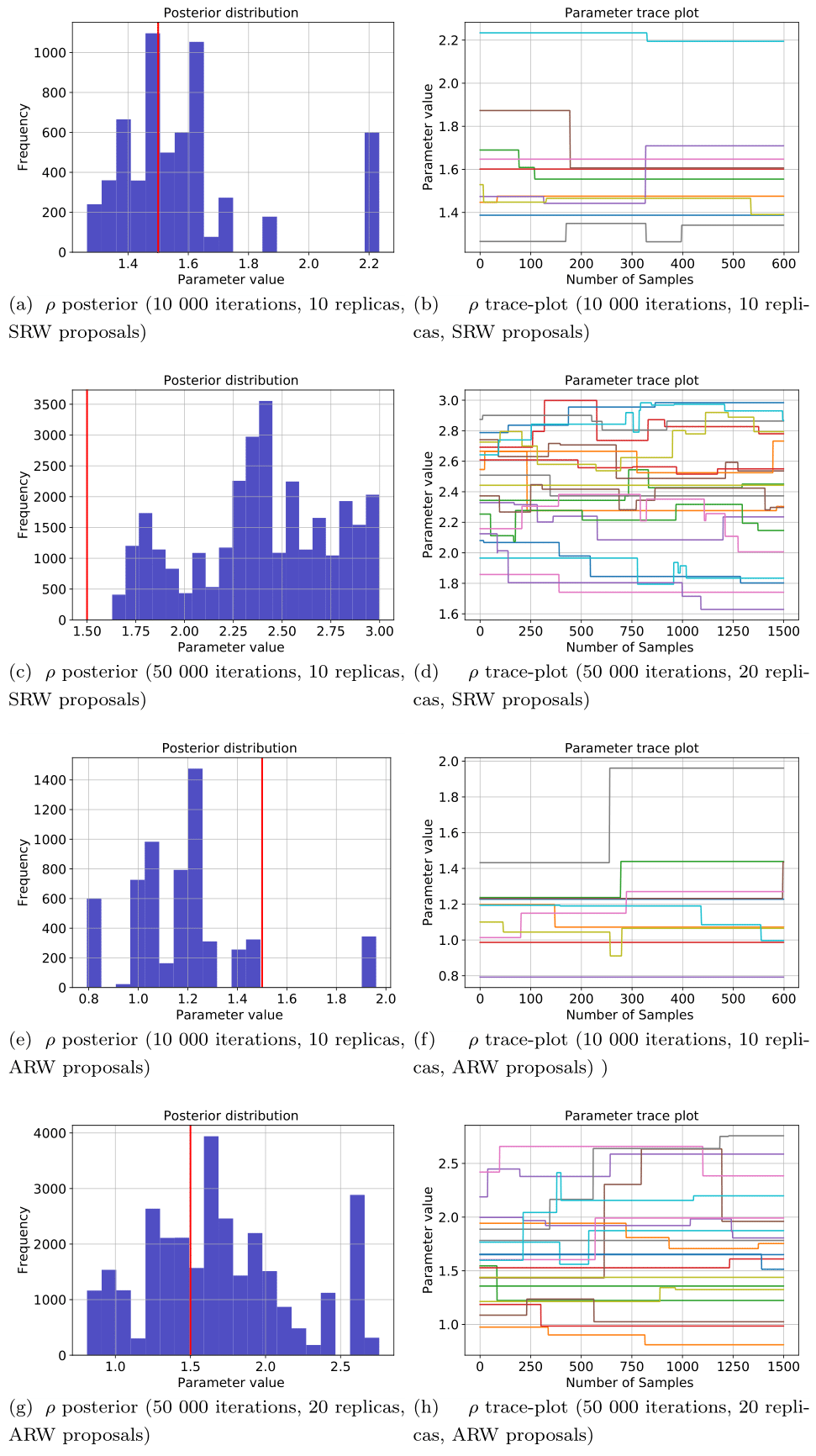}
    
    \caption{  \textcolor{black}{  Posterior distribution and trace-plot during sampling for precipitation ($\rho$) parameter  for different sampling times for the Cr problem using SRW and ARW proposal distributions in PT-Bayeslands. Panels (a)~and~(b) correspond to 10000 iterations with a SRW proposal. Panels (c)~and~(d) correspond to 50000 iterations with a SRW proposal. Panels (e)~and~(f) correspond to 10000 iterations with an ARW proposal and panels~(g)~and (h) correspond to 50000 iterations with an ARW proposal. } }
 \label{fig:crater_pos}
  \end{center}
\end{figure}

\begin{figure}[htb!]
  \begin{center}
    \includegraphics[width=110mm]{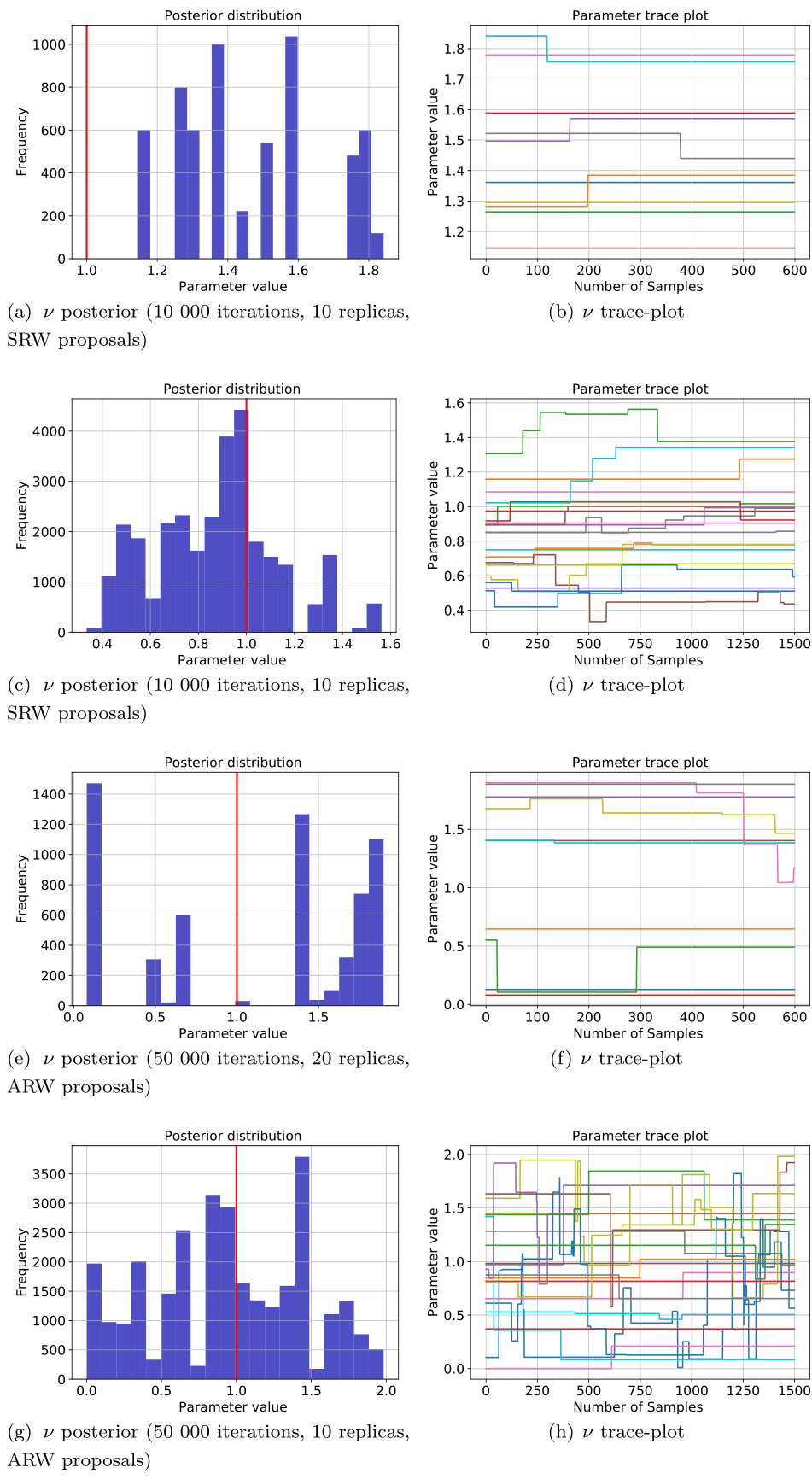}

    \caption{   \textcolor{black}{  Posterior distribution and trace-plot during sampling for n-value  ($\nu$) parameter for different sampling times for the Cr problem using SRW and ARW proposal distributions in PT-Bayeslands. Panels (a)~and~(b) correspond to 10000 iterations with a SRW proposal. Panels (c)~and~(d) correspond to 50000 iterations with a SRW proposal. Panels (e)~and~(f) correspond to 10 000 iterations with an ARW proposal and panels~(g)~and (h) correspond to 50 000 iterations with an ARW proposal. }   }
 \label{fig:cm_pos}
  \end{center}
\end{figure}

\begin{figure}[htb!]
  \begin{center}
    \includegraphics[width=110mm]{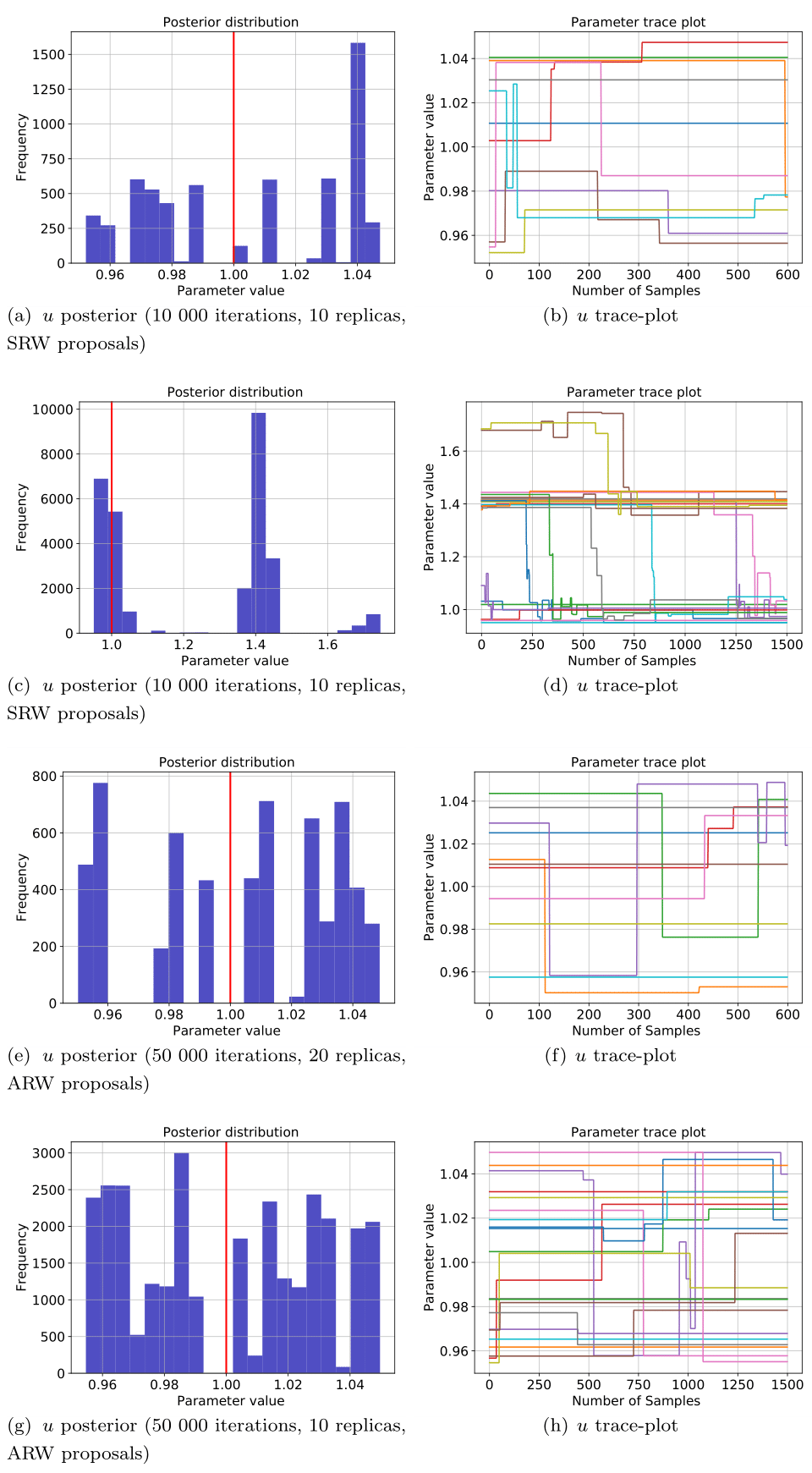}

    \caption{   \textcolor{black}{  Posterior distribution and trace-plot during sampling for uplift ($u$) parameter  for different sampling times for the Cr problem using SRW and ARW proposal distributions in PT-Bayeslands. Panels (a)~and~(b) correspond to 10000 iterations with a SRW proposal. Panels (c)~and~(d) correspond to 50000 iterations with a SRW proposal. Panels (e)~and~(f) correspond to 10 000 iterations with an ARW proposal and panels~(g)~and (h) correspond to 50 000 iterations with an ARW proposal. }   }
 \label{fig:mountain_pos}
  \end{center}
\end{figure}

 We present further details for selected parameters for the given problems. We show details of the precipitation posterior for Cr and CM problems. We then show the Mt model uplift posterior as this is a unique parameter not present in the other problems. Figures \ref{fig:crater_pos}, \ref{fig:mountain_pos} and \ref{fig:cm_pos} show the posterior distribution and trace-plot for selected parameters for the respective problems.   Note that the most extensive parameter space exploration by PT-Bayeslands occurs for the Cr problem. The exploration ability of the replicas is visible, given the trace-plot in Figure \ref{fig:crater_pos}. This is in contrast with the exploration shown in trace-plot given in Mt and CM problems (\ref{fig:mountain_pos} and \ref{fig:cm_pos}), since they have a highly irregular log-likelihood surface which has a number of sub-optimal peaks. The trace-plot shows that the replicas are essentially trapped in these sub-optimal peaks; increasing the number of replicas (from 10 to 20) and sampling time ( from 10,000 to 50 0,000), shows no major difference. However, even these sub-optimal peaks give an acceptable prediction as shown in  Figure \ref{fig:cross_section}.

In this section, we only presented selected details of the posterior distribution, successive topography, and successive erosion-deposition predictions. Detailed results for all problems are provided in 
supplementary material online \footnote{Supplementary results: 
\url{
https://github.com/badlands-model/paralleltemp_Bayeslands/tree/master/supplementary_results}}.

 \section{Discussion}
 
The results presented in the previous section  can be summarised as follows; \begin{enumerate}
    \item For multi-modal problems PT-Bayeslands provides better exploration of the  
posterior than SC-Bayeslands. This  improvement in performance is particularly visible when the modes are not connected (as in the CM  Mt problem) and as the dimension of the problem increases.
Despite 
the variability in results between SC-Bayeslands and PT-Bayeslands, the predictive performance for both methods is similar. This  suggests that  SC-Bayeslands has the ability to converge in a sub-optimal mode, even if it does not explore the multi-modal  posterior fully. These are important results because multi-modal posterior distributions are typically present in geological and geophysical inversion problems
\citep{dalla2015challenges,beskos2017}. We observed that  different experimental runs, depending on initial conditions (starting value), only certain combination of the modes can be recovered.

\item \textcolor{black}{Sampling multimodal distribution  is a challenge; however, our results show that   convergence on sub-optimal modes gives similar topography evolution when compared with true values. The use of better proposal distributions can help in future work, such as using gradient-free  meta-heuristics  from field of evolutionary algorithms \citep{
drugan2003evolutionary,ter2006markov}. Moreover, we also need to incorporate  additional data or model constraints that could better  help reduce the flexibility of the solution.}

\item
The  computation time for a given number of iterations 
does not scale linearly with the number of replicas.  There is a trade-off 
between proposing to swap chains and the rate of convergence of the 
PT-Bayeslands. Swapping proposals between neighboring replicas 
gives better mixing but introduces computational overhead in a parallel computing environment since each replica needs 
to wait for all   to complete sampling until their  swap-interval in order to compute the neighbor swap 
probability by the managing process. This computational overhead can be  higher if the number of replicas is 
large. We note that the synthetic problems considered in this paper, only 
required a few seconds of simulation time for Badlands, whereas, in real-world 
problems, each simulation of Badlands could take several minutes to hours. In 
such cases, the trade-off between the frequency of swapping and prediction 
accuracy would need to be evaluated.
\item 
Increasing 
the number of cores does not necessarily mean that the prediction accuracy will 
get better. In particular, it is advantageous to use a larger number of cores for 
large scale problems where the Badlands model takes hours to evaluate a single 
proposal.
\item The ARW and SRW proposal distributions have the same predictive accuracy; however, the ARW within chain proposal distribution explores the posterior better than the SRW proposal which is  evident by the convergence diagnosis shown (Table \ref{tab:convergence}).

\item \textcolor{black}{We highlight that a major limitation in the experiments  is the assumption that initial topographies (paleo-elevation) is easily available. We simulated the initial topography and demonstrated the framework with small-scale  synthetic problems. We need to note that initial topography   is difficult to obtain given limited observations and hence it is a challenge to reconstruct initial topography. The estimation of topography regions which data is not available can be tackled by Bayeslands in future work.  The framework can be used to fuse multiple sources of information about initial topography which includes, expert knowledge, data and models for paleo-elevation.}

\item     \textcolor{black}{ The issue of multimodality can be understood by the interpretation of the different factors that affect the topography development over time.  Further work could also consider innovative ways to capture these processes in the likelihood function, such as  paleo ground-truth data about ancient streams and rivers. Such data can help further constrain the sampling and address problems that arise from multimodlaity. }

\item  \textcolor{black}{In Badlands, similar to the model from\citep{croissant2014constraining}, we compute the erosion using a stream power law that links the erosion rate to both the erodibility coefficient of soil, the drainage area and the slope. In future work, the Bayeslands framework  could  consider other related models. However,  changes to the landscape evolution model would feature additional parameters and could have a slightly different likelihood function in order  to incorporate the streams and drainage rate that would affect sedimentary deposits. }

\end{enumerate}

 \section{Conclusions and Future Work}
 
 \textcolor{black}{ The PT-Bayeslands  framework  provides a   rigorous approach for estimation and uncertainty quantification of free parameters in Badlands landscape evolution model  using  multi-core parallel 
tempering. We provide  a comprehensive experimental evaluation of the method given different combinations of user settings that include sampling time,  number of replicas, swap-rate for different types of problems. We also provided a comparison of two different proposal distributions with convergence diagnosis. In general,  the  results show that the method not only reduces the computation time, but also provides a means to explore the 
parameter space  in highly irregular  multi-modal distributions. This has been demonstrated by the results that show better posterior distributions of the parameters along with improvement in prediction accuracy of  topography and sediment  deposition. }
 
 There are several areas of future work. The first is the impact of the temperature ladder on convergence of the MCMC scheme for geophysical inversion problems, see for example \citep{patriksson2008temperature}.  The second is to allow  for some  parameters, such as precipitation, to vary across space and time. Such a model would allow us to measure the impact of climate change on the environment in terms of sediment erosion/deposition and elevation. We further note that the proposed framework is general and hence, apart from Badlands,  it can accommodate other landscape evolution models \citep{coulthard2001landscape} such as the Landlab model \citep{hobley2017creative}.
 

The approach outlined here could be extended by surrogate-assisted models where a surrogate of Badlands would be implemented via machine learning, evaluating the proposals in a fraction of the time taken by actual Badlands runs. This would be of interest for landscape and basin evolution models over long geological time intervals (100 million years and longer) at high resolution whose run time would be prohibitive for a PT-Bayeslands application as outlined here. Further, efficient gradient free proposals would need to be constructed as the number 
of parameters and the complexity of the model increases. 

\acknowledgments

We acknowledge the Strategic Research Excellence Initiative (SREI) Grant from the University of 
Sydney. Furthermore, we  acknowledge the Artemis high performance 
computing infrastructure provided by Sydney Informatics Hub of the University 
of Sydney.  This work was supported by Australian Research Council grant IH130200012. We  sincerely thank Dr. Richard Scalzo for valuable discussions and  Ashray Aman for technical support. Furthermore, we would like to sincerely thank the anonymous reviewers and the handling editor for helping us improve the paper. 
\section*{Data and supplementary material }
  
  \textcolor{black}{
The selected landscape evolution problems been adapted from  examples of the  Badlands model \footnote{\url{https://github.com/badlands-model/pyBadlands/tree/master/Examples}}.} \textcolor{black}{
Additional results with open software Python  code and data is available online  \footnote{\url{https://github.com/intelligentEarth/pt-Bayeslands}}. } Supplementary  results for the respective  problems is also  provided  online \footnote{ 
\url{
https://github.com/badlands-model/paralleltemp_Bayeslands/tree/master/supplementary_results}}.

 \section*{Appendix}

\begin{figure}[htb!]
  \begin{center}
    \includegraphics[width=90mm]{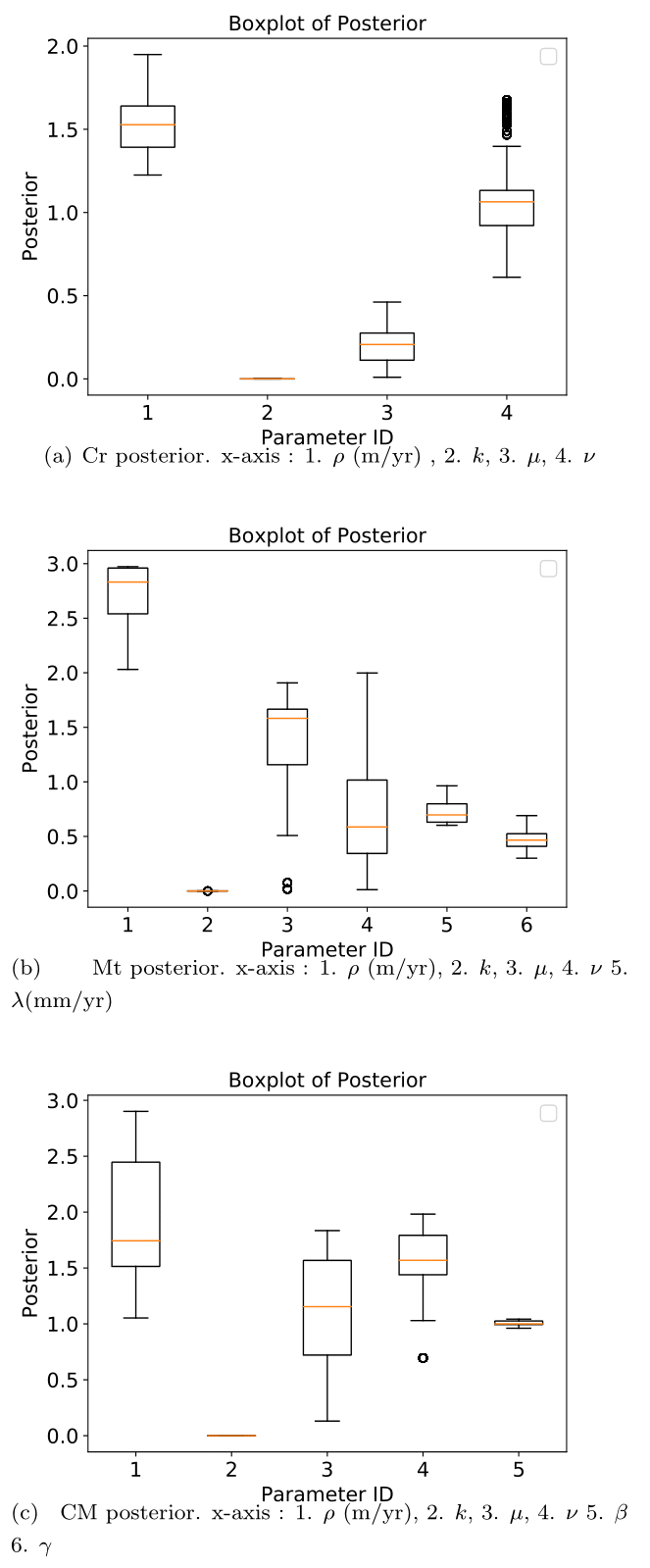}
    
    \caption{Boxplot showing the posterior distribution of the free parameters for the respective problems. }
 \label{fig:boxplot}
  \end{center}
\end{figure}

\begin{figure}[htb!]
  \begin{center}
    \includegraphics[width=90mm]{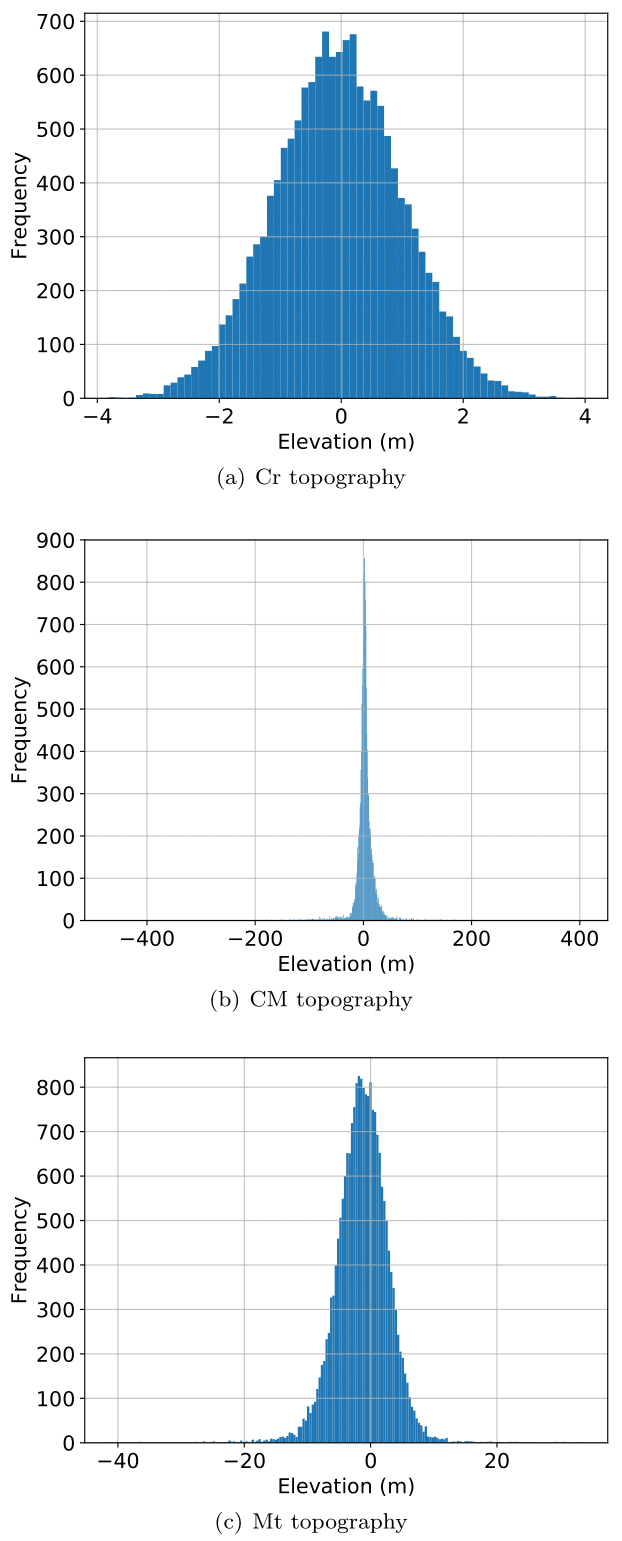}
    
    \caption{Distribution of topography residuals for the respective problems (Cr, Mt and CM) }
 \label{fig:residuals}
  \end{center}
\end{figure}

 Boxplot showing the posterior distribution of the free parameters for the respective problems (Figure \ref{fig:boxplot}). The residuals for each topography is shown in Figure \ref{fig:residuals}.

\bibliographystyle{agufull08}
\bibliography{sample}

\begin{thebibliography}{72}
\providecommand{\natexlab}[1]{#1}
\expandafter\ifx\csname urlstyle\endcsname\relax
  \providecommand{\doi}[1]{doi:\discretionary{}{}{}#1}\else
  \providecommand{\doi}{doi:\discretionary{}{}{}\begingroup
  \urlstyle{rm}\Url}\fi

\bibitem[{\textit{Adams et~al.}(2017)\textit{Adams, Gasparini, Hobley, Tucker,
  Hutton, Nudurupati, and Istanbulluoglu}}]{Adams2017}
Adams, J.~M., N.~M. Gasparini, D.~E.~J. Hobley, G.~E. Tucker, E.~W.~H. Hutton,
  S.~S. Nudurupati, and E.~Istanbulluoglu (2017), The landlab v1.0 overlandflow
  component: a python tool for computing shallow-water flow across watersheds,
  \textit{Geoscientific Model Development}, \textit{10}(4), 1645--1663.

\bibitem[{\textit{Beskos et~al.}(2017)\textit{Beskos, Girolami, Lan, Farrell,
  and Stuart}}]{beskos2017}
Beskos, A., M.~Girolami, S.~Lan, P.~E. Farrell, and A.~M. Stuart (2017),
  Geometric mcmc for infinite-dimensional inverse problems, \textit{Journal of
  Computational Physics}, \textit{335}, 327--351.

\bibitem[{\textit{Campforts et~al.}(2017)\textit{Campforts, Schwanghart, and
  Govers}}]{Campforts2017}
Campforts, B., W.~Schwanghart, and G.~Govers (2017), Accurate simulation of
  transient landscape evolution by eliminating numerical diffusion: the
  ttlem~1.0 model, \textit{Earth Surface Dynamics}, \textit{5}(1), 47--66.

\bibitem[{\textit{Chandra et~al.}(2019{\natexlab{a}})\textit{Chandra, Jain,
  Deo, and Cripps}}]{CHANDRA_NC2019}
Chandra, R., K.~Jain, R.~V. Deo, and S.~Cripps (2019{\natexlab{a}}),
  Langevin-gradient parallel tempering for bayesian neural learning,
  \textit{Neurocomputing}, \doi{https://doi.org/10.1016/j.neucom.2019.05.082}.

\bibitem[{\textit{Chandra et~al.}(2019{\natexlab{b}})\textit{Chandra, Azam,
  Müller, Salles, and Cripps}}]{CHANDRA2019}
Chandra, R., D.~Azam, R.~D. Müller, T.~Salles, and S.~Cripps
  (2019{\natexlab{b}}), Bayeslands: A bayesian inference approach for parameter
  uncertainty quantification in badlands, \textit{Computers \& Geosciences},
  \doi{https://doi.org/10.1016/j.cageo.2019.06.012}.

\bibitem[{\textit{Chen et~al.}(2014)\textit{Chen, Darbon, and Morel}}]{Chen14}
Chen, A., J.~Darbon, and J.-M. Morel (2014), {Landscape evolution models: a
  review of their fundamental equations.}, \textit{Geomorphology},
  \textit{219}, 68–86.

\bibitem[{\textit{Chib and Greenberg}(1995)}]{chib1995understanding}
Chib, S., and E.~Greenberg (1995), Understanding the metropolis-hastings
  algorithm, \textit{The american statistician}, \textit{49}(4), 327--335.

\bibitem[{\textit{Coulthard}(2001)}]{coulthard2001landscape}
Coulthard, T.~J. (2001), Landscape evolution models: a software review,
  \textit{Hydrological processes}, \textit{15}(1), 165--173.

\bibitem[{\textit{Croissant and Braun}(2014)}]{croissant2014constraining}
Croissant, T., and J.~Braun (2014), Constraining the stream power law: a novel
  approach combining a landscape evolution model and an inversion method.,
  \textit{Earth surface dynamics.}, \textit{2}(1), 155--166.

\bibitem[{\textit{Dalla~Mura et~al.}(2015)\textit{Dalla~Mura, Prasad, Pacifici,
  Gamba, Chanussot, and Benediktsson}}]{dalla2015challenges}
Dalla~Mura, M., S.~Prasad, F.~Pacifici, P.~Gamba, J.~Chanussot, and J.~A.
  Benediktsson (2015), Challenges and opportunities of multimodality and data
  fusion in remote sensing, \textit{Proceedings of the IEEE}, \textit{103}(9),
  1585--1601.

\bibitem[{\textit{Drugan and Thierens}(2003)}]{drugan2003evolutionary}
Drugan, M.~M., and D.~Thierens (2003), Evolutionary markov chain monte carlo,
  in \textit{International Conference on Artificial Evolution (Evolution
  Artificielle)}, pp. 63--76, Springer.

\bibitem[{\textit{Fox et~al.}(2014)\textit{Fox, Reverman, Herman, Fellin,
  Sternai, and Willett}}]{fox2014rock}
Fox, M., R.~Reverman, F.~Herman, M.~G. Fellin, P.~Sternai, and S.~D. Willett
  (2014), Rock uplift and erosion rate history of the bergell intrusion from
  the inversion of low temperature thermochronometric data,
  \textit{Geochemistry, Geophysics, Geosystems}, \textit{15}(4), 1235--1257.

\bibitem[{\textit{Fox et~al.}(2015)\textit{Fox, Bodin, and
  Shuster}}]{fox2015abrupt}
Fox, M., T.~Bodin, and D.~L. Shuster (2015), Abrupt changes in the rate of
  andean plateau uplift from reversible jump markov chain monte carlo inversion
  of river profiles, \textit{Geomorphology}, \textit{238}, 1--14.

\bibitem[{\textit{Gallagher et~al.}(2009)\textit{Gallagher, Charvin, Nielsen,
  Sambridge, and Stephenson}}]{gallagher2009markov}
Gallagher, K., K.~Charvin, S.~Nielsen, M.~Sambridge, and J.~Stephenson (2009),
  Markov chain monte carlo (mcmc) sampling methods to determine optimal models,
  model resolution and model choice for earth science problems, \textit{Marine
  and Petroleum Geology}, \textit{26}(4), 525--535.

\bibitem[{\textit{Gasparini and Brandon}(2011)}]{gasparini2011generalized}
Gasparini, N.~M., and M.~T. Brandon (2011), A generalized power law
  approximation for fluvial incision of bedrock channels, \textit{Journal of
  Geophysical Research: Earth Surface}, \textit{116}(F2).

\bibitem[{\textit{Gelman et~al.}(1992)\textit{Gelman, Rubin
  et~al.}}]{gelman1992inference}
Gelman, A., D.~B. Rubin, et~al. (1992), Inference from iterative simulation
  using multiple sequences, \textit{Statistical science}, \textit{7}(4),
  457--472.

\bibitem[{\textit{Geyer and Thompson}(1995)}]{geyer1995annealing}
Geyer, C.~J., and E.~A. Thompson (1995), Annealing markov chain monte carlo
  with applications to ancestral inference, \textit{Journal of the American
  Statistical Association}, \textit{90}(431), 909--920.

\bibitem[{\textit{Girolami and Calderhead}(2011)}]{girolami2011riemann}
Girolami, M., and B.~Calderhead (2011), Riemann manifold langevin and
  hamiltonian monte carlo methods, \textit{Journal of the Royal Statistical
  Society: Series B (Statistical Methodology)}, \textit{73}(2), 123--214.

\bibitem[{\textit{Godard et~al.}(2006)\textit{Godard, Lav{\'e}, and
  Cattin}}]{godard2006numerical}
Godard, V., J.~Lav{\'e}, and R.~Cattin (2006), Numerical modelling of erosion
  processes in the himalayas of nepal: Effects of spatial variations of rock
  strength and precipitation, \textit{Geological Society, London, Special
  Publications}, \textit{253}(1), 341--358.

\bibitem[{\textit{Goren et~al.}(2014)\textit{Goren, Fox, and
  Willett}}]{goren2014tectonics}
Goren, L., M.~Fox, and S.~D. Willett (2014), Tectonics from fluvial topography
  using formal linear inversion: Theory and applications to the inyo mountains,
  california, \textit{Journal of Geophysical Research: Earth Surface},
  \textit{119}(8), 1651--1681.

\bibitem[{\textit{Grandis et~al.}(1999)\textit{Grandis, Menvielle, and
  Roussignol}}]{grandis1999bayesian}
Grandis, H., M.~Menvielle, and M.~Roussignol (1999), Bayesian inversion with
  markov chains—i. the magnetotelluric one-dimensional case,
  \textit{Geophysical Journal International}, \textit{138}(3), 757--768.

\bibitem[{\textit{Haario et~al.}(2001)\textit{Haario, Saksman, Tamminen
  et~al.}}]{haario2001adaptive}
Haario, H., E.~Saksman, J.~Tamminen, et~al. (2001), An adaptive metropolis
  algorithm, \textit{Bernoulli}, \textit{7}(2), 223--242.

\bibitem[{\textit{Harel et~al.}(2016)\textit{Harel, Mudd, and
  Attal}}]{HAREL2016184}
Harel, M.-A., S.~Mudd, and M.~Attal (2016), Global analysis of the stream power
  law parameters based on worldwide 10be denudation rates,
  \textit{Geomorphology}, \textit{268}, 184 -- 196.

\bibitem[{\textit{Hastings}(1970)}]{hastings1970monte}
Hastings, W.~K. (1970), Monte carlo sampling methods using markov chains and
  their applications, \textit{Biometrika}, \textit{57}(1), 97--109.

\bibitem[{\textit{Hobley et~al.}(2017)\textit{Hobley, Adams, Nudurupati,
  Hutton, Gasparini, Istanbulluoglu, and Tucker}}]{hobley2017creative}
Hobley, D.~E., J.~M. Adams, S.~S. Nudurupati, E.~W. Hutton, N.~M. Gasparini,
  E.~Istanbulluoglu, and G.~E. Tucker (2017), Creative computing with landlab:
  an open-source toolkit for building, coupling, and exploring two-dimensional
  numerical models of earth-surface dynamics, \textit{Earth Surface Dynamics},
  \textit{5}(1), 21.

\bibitem[{\textit{Hobley et~al.}(2011)\textit{Hobley, Sinclair, Mudd, and
  Cowie}}]{Hobley2011}
Hobley, D. E.~J., H.~D. Sinclair, S.~M. Mudd, and P.~A. Cowie (2011), Field
  calibration of sediment flux dependent river incision, \textit{Journal of
  Geophysical Research: Earth Surface}, \textit{116}(F4).

\bibitem[{\textit{Hoffman and Gelman}(2014)}]{hoffman2014no}
Hoffman, M.~D., and A.~Gelman (2014), The no-u-turn sampler: adaptively setting
  path lengths in hamiltonian monte carlo., \textit{Journal of Machine Learning
  Research}, \textit{15}(1), 1593--1623.

\bibitem[{\textit{Howard et~al.}(1994)\textit{Howard, Dietrich, and
  Seidl}}]{Howard1994}
Howard, A.~D., W.~E. Dietrich, and M.~A. Seidl (1994), Modeling fluvial erosion
  on regional to continental scales, \textit{Journal of Geophysical Research:
  Solid Earth}, \textit{99}(B7), 13,971--13,986.

\bibitem[{\textit{Karimi et~al.}(2011)\textit{Karimi, Dickson, and
  Hamze}}]{karimi2011high}
Karimi, K., N.~Dickson, and F.~Hamze (2011), High-performance physics
  simulations using multi-core cpus and gpgpus in a volunteer computing
  context, \textit{The International Journal of High Performance Computing
  Applications}, \textit{25}(1), 61--69.

\bibitem[{\textit{Kone and Kofke}(2005)}]{kone2005selection}
Kone, A., and D.~A. Kofke (2005), Selection of temperature intervals for
  parallel-tempering simulations, \textit{The Journal of chemical physics},
  \textit{122}(20), 206,101.

\bibitem[{\textit{Lamport}(1986)}]{lamport1986interprocess}
Lamport, L. (1986), On interprocess communication, \textit{Distributed
  computing}, \textit{1}(2), 86--101.

\bibitem[{\textit{Li et~al.}(2009)\textit{Li, Mascagni, and Gorin}}]{LI2009269}
Li, Y., M.~Mascagni, and A.~Gorin (2009), A decentralized parallel
  implementation for parallel tempering algorithm, \textit{Parallel Computing},
  \textit{35}(5), 269 -- 283.

\bibitem[{\textit{Malinverno}(2002)}]{malinverno2002parsimonious}
Malinverno, A. (2002), Parsimonious bayesian markov chain monte carlo inversion
  in a nonlinear geophysical problem, \textit{Geophysical Journal
  International}, \textit{151}(3), 675--688.

\bibitem[{\textit{Maraschini and Foti}(2010)}]{maraschini2010monte}
Maraschini, M., and S.~Foti (2010), A monte carlo multimodal inversion of
  surface waves, \textit{Geophysical Journal International}, \textit{182}(3),
  1557--1566.

\bibitem[{\textit{Marinari and Parisi}(1992)}]{marinari1992}
Marinari, E., and G.~Parisi (1992), Simulated tempering: a new monte carlo
  scheme, \textit{EPL (Europhysics Letters)}, \textit{19}(6), 451.

\bibitem[{\textit{Metropolis et~al.}(1953)\textit{Metropolis, Rosenbluth,
  Rosenbluth, Teller, and Teller}}]{metropolis1953equation}
Metropolis, N., A.~W. Rosenbluth, M.~N. Rosenbluth, A.~H. Teller, and E.~Teller
  (1953), Equation of state calculations by fast computing machines,
  \textit{The journal of chemical physics}, \textit{21}(6), 1087--1092.

\bibitem[{\textit{Miasojedow et~al.}(2013)\textit{Miasojedow, Moulines, and
  Vihola}}]{miasojedow2013adaptive}
Miasojedow, B., E.~Moulines, and M.~Vihola (2013), An adaptive parallel
  tempering algorithm, \textit{Journal of Computational and Graphical
  Statistics}, \textit{22}(3), 649--664.

\bibitem[{\textit{Mills et~al.}(1992)\textit{Mills, Fox, and
  Heimbach}}]{mills1992implementing}
Mills, K., G.~Fox, and R.~Heimbach (1992), Implementing an intervisibility
  analysis model on a parallel computing system, \textit{Computers \&
  Geosciences}, \textit{18}(8), 1047--1054.

\bibitem[{\textit{Mingas et~al.}(2017)\textit{Mingas, Bottolo, and
  Bouganis}}]{mingas2017particle}
Mingas, G., L.~Bottolo, and C.-S. Bouganis (2017), Particle mcmc algorithms and
  architectures for accelerating inference in state-space models,
  \textit{International Journal of Approximate Reasoning}, \textit{83},
  413--433.

\bibitem[{\textit{Mosegaard and Tarantola}(1995)}]{mosegaard1995monte}
Mosegaard, K., and A.~Tarantola (1995), Monte carlo sampling of solutions to
  inverse problems, \textit{Journal of Geophysical Research: Solid Earth},
  \textit{100}(B7), 12,431--12,447.

\bibitem[{\textit{Mosegaard and Vestergaard}(1991)}]{mosegaard1991simulated}
Mosegaard, K., and P.~D. Vestergaard (1991), A simulated annealing approach to
  seismic model optimization with sparse prior information, \textit{Geophysical
  Prospecting}, \textit{39}(5), 599--611.

\bibitem[{\textit{Neal}(1996)}]{neal1996sampling}
Neal, R.~M. (1996), Sampling from multimodal distributions using tempered
  transitions, \textit{Statistics and computing}, \textit{6}(4), 353--366.

\bibitem[{\textit{Neal et~al.}(2011)}]{neal2011mcmc}
Neal, R.~M., et~al. (2011), Mcmc using hamiltonian dynamics, \textit{Handbook
  of Markov Chain Monte Carlo}, \textit{2}(11).

\bibitem[{\textit{Patriksson and van~der
  Spoel}(2008)}]{patriksson2008temperature}
Patriksson, A., and D.~van~der Spoel (2008), A temperature predictor for
  parallel tempering simulations, \textit{Physical Chemistry Chemical Physics},
  \textit{10}(15), 2073--2077.

\bibitem[{\textit{Reid et~al.}(2013)\textit{Reid, Bonilla, McCalman, Rawling,
  and Ramos}}]{reid2013bayesian}
Reid, A., E.~V. Bonilla, L.~McCalman, T.~Rawling, and F.~Ramos (2013), Bayesian
  joint inversions for the exploration of {Earth} resources, in
  \textit{{IJCAI}}, pp. 2877--2884.

\bibitem[{\textit{Rejman et~al.}(1998)\textit{Rejman, Turski, and
  Paluszek}}]{rejman1998spatial}
Rejman, J., R.~Turski, and J.~Paluszek (1998), Spatial and temporal variations
  in erodibility of loess soil, \textit{Soil and Tillage Research},
  \textit{46}(1-2), 61--68.

\bibitem[{\textit{Robert and Casella}(2011)}]{robert2011short}
Robert, C., and G.~Casella (2011), A short history of markov chain monte carlo:
  Subjective recollections from incomplete data, \textit{Statistical Science},
  pp. 102--115.

\bibitem[{\textit{Roberts and White}(2010)}]{roberts2010estimating}
Roberts, G.~G., and N.~White (2010), Estimating uplift rate histories from
  river profiles using african examples, \textit{Journal of Geophysical
  Research: Solid Earth}, \textit{115}(B2).

\bibitem[{\textit{Rocca et~al.}(2009)\textit{Rocca, Benedetti, Donelli,
  Franceschini, and Massa}}]{rocca2009evolutionary}
Rocca, P., M.~Benedetti, M.~Donelli, D.~Franceschini, and A.~Massa (2009),
  Evolutionary optimization as applied to inverse scattering problems,
  \textit{Inverse Problems}, \textit{25}(12), 123,003.

\bibitem[{\textit{Salinger}(1980)}]{salinger1980new}
Salinger, M. (1980), New zealand climate: I. precipitation patterns,
  \textit{Monthly weather review}, \textit{108}(11), 1892--1904.

\bibitem[{\textit{Salles}(2016)}]{salles2016b}
Salles, T. (2016), Badlands: A parallel basin and landscape dynamics model,
  \textit{SoftwareX}, \textit{5}, 195--202.

\bibitem[{\textit{Salles and Duclaux}(2015)}]{Salles15}
Salles, T., and G.~Duclaux (2015), {Combined hillslope diffusion and sediment
  transport simulation applied to landscape dynamics modelling.}, \textit{Earth
  Surf. Process Landf.}, \textit{40}(6), 823–39.

\bibitem[{\textit{Salles and Hardiman}(2016)}]{salles2016badlands}
Salles, T., and L.~Hardiman (2016), Badlands: An open-source, flexible and
  parallel framework to study landscape dynamics, \textit{Computers \&
  Geosciences}, \textit{91}, 77--89.

\bibitem[{\textit{Salles et~al.}(2017)\textit{Salles, Flament, and
  M{\"u}ller}}]{salles2017influence}
Salles, T., N.~Flament, and D.~M{\"u}ller (2017), Influence of mantle flow on
  the drainage of eastern australia since the jurassic period,
  \textit{Geochemistry, Geophysics, Geosystems}, \textit{18}(1), 280--305.

\bibitem[{\textit{Salles et~al.}(2018{\natexlab{a}})\textit{Salles, Ding, and
  Brocard}}]{salles2018pybadlands}
Salles, T., X.~Ding, and G.~Brocard (2018{\natexlab{a}}), pybadlands: A
  framework to simulate sediment transport, landscape dynamics and basin
  stratigraphic evolution through space and time, \textit{PloS one},
  \textit{13}(4), e0195,557.

\bibitem[{\textit{Salles et~al.}(2018{\natexlab{b}})\textit{Salles, Ding, and
  Brocard}}]{salles2018a}
Salles, T., X.~Ding, and G.~Brocard (2018{\natexlab{b}}), pybadlands: A
  framework to simulate sediment transport, landscape dynamics and basin
  stratigraphic evolution through space and time, \textit{PLOS ONE},
  \textit{13}, 1--24.

\bibitem[{\textit{Salles et~al.}(2018{\natexlab{c}})\textit{Salles, Ding,
  Webster, Vila-Concejo, Brocard, and Pall}}]{salles2018b}
Salles, T., X.~Ding, J.~M. Webster, A.~Vila-Concejo, G.~Brocard, and J.~Pall
  (2018{\natexlab{c}}), A unified framework for modelling sediment fate from
  source to sink and its interactions with reef systems over geological times,
  \textit{Scientific Reports}, \textit{8}(1), 5252,
  \doi{10.1038/s41598-018-23519-8}.

\bibitem[{\textit{Sambridge}(1999)}]{sambridge1999geophysical}
Sambridge, M. (1999), Geophysical inversion with a neighbourhood
  algorithm—ii. appraising the ensemble, \textit{Geophysical Journal
  International}, \textit{138}(3), 727--746.

\bibitem[{\textit{Sambridge}(2013)}]{sambridge2013parallel}
Sambridge, M. (2013), A parallel tempering algorithm for probabilistic sampling
  and multimodal optimization, \textit{Geophysical Journal International},
  \textit{196}(1), 357--374.

\bibitem[{\textit{Sambridge and Mosegaard}(2002)}]{sambridge2002monte}
Sambridge, M., and K.~Mosegaard (2002), Monte carlo methods in geophysical
  inverse problems, \textit{Reviews of Geophysics}, \textit{40}(3).

\bibitem[{\textit{Scalzo et~al.}(2019)\textit{Scalzo, Kohn, Olierook, Houseman,
  Chandra, Girolami, and Cripps}}]{Scalzo2018GMD}
Scalzo, R., D.~Kohn, H.~Olierook, G.~Houseman, R.~Chandra, M.~Girolami, and
  S.~Cripps (2019), Efficiency and robustness in monte carlo sampling for 3-d
  geophysical inversions with obsidian v0.1.2: setting up for success,
  \textit{Geoscientific Model Development}, \textit{12}(7), 2941--2960,
  \doi{10.5194/gmd-12-2941-2019}.

\bibitem[{\textit{Scott and Zhang}(1990)}]{scott1990finite}
Scott, L.~R., and S.~Zhang (1990), Finite element interpolation of nonsmooth
  functions satisfying boundary conditions, \textit{Mathematics of
  Computation}, \textit{54}(190), 483--493.

\bibitem[{\textit{Sen and Stoffa}(1996)}]{sen1996bayesian}
Sen, M.~K., and P.~L. Stoffa (1996), Bayesian inference, gibbs' sampler and
  uncertainty estimation in geophysical inversion, \textit{Geophysical
  Prospecting}, \textit{44}(2), 313--350.

\bibitem[{\textit{Sen and Stoffa}(2013)}]{sen2013global}
Sen, M.~K., and P.~L. Stoffa (2013), \textit{Global optimization methods in
  geophysical inversion}, Cambridge University Press.

\bibitem[{\textit{Stock and Montgomery}(1999)}]{stock1999geologic}
Stock, J.~D., and D.~R. Montgomery (1999), Geologic constraints on bedrock
  river incision using the stream power law, \textit{Journal of Geophysical
  Research: Solid Earth}, \textit{104}(B3), 4983--4993.

\bibitem[{\textit{Tarantola}(2006)}]{tarantola2006popper}
Tarantola, A. (2006), Popper, bayes and the inverse problem, \textit{Nature
  physics}, \textit{2}(8), 492.

\bibitem[{\textit{Ter~Braak}(2006)}]{ter2006markov}
Ter~Braak, C.~J. (2006), A markov chain monte carlo version of the genetic
  algorithm differential evolution: easy bayesian computing for real parameter
  spaces, \textit{Statistics and Computing}, \textit{16}(3), 239--249.

\bibitem[{\textit{Tucker and Hancock}(2010)}]{Tucker10}
Tucker, G.~E., and G.~R. Hancock (2010), Modelling landscape evolution,
  \textit{Earth Surface Processes and Landforms}, \textit{35}(1), 28--50.

\bibitem[{\textit{Ummenhofer and England}(2007)}]{ummenhofer2007interannual}
Ummenhofer, C.~C., and M.~H. England (2007), Interannual extremes in new
  zealand precipitation linked to modes of southern hemisphere climate
  variability, \textit{Journal of Climate}, \textit{20}(21), 5418--5440.

\bibitem[{\textit{Vrugt et~al.}(2006)\textit{Vrugt, Nuall{\'a}in, Robinson,
  Bouten, Dekker, and Sloot}}]{vrugt2006application}
Vrugt, J.~A., B.~O. Nuall{\'a}in, B.~A. Robinson, W.~Bouten, S.~C. Dekker, and
  P.~M. Sloot (2006), Application of parallel computing to stochastic parameter
  estimation in environmental models, \textit{Computers \& Geosciences},
  \textit{32}(8), 1139--1155.

\bibitem[{\textit{Whipple and Tucker}(2002)}]{Whipple2002}
Whipple, K.~X., and G.~E. Tucker (2002), Implications of
  sediment-flux-dependent river incision models for landscape evolution,
  \textit{Journal of Geophysical Research: Solid Earth}, \textit{107}(B2),
  1--20.

\bibitem[{\textit{Zhang et~al.}(2007)\textit{Zhang, Liu, Shi, Yuen, Yan, and
  Liang}}]{zhang2007toward}
Zhang, H., M.~Liu, Y.~Shi, D.~A. Yuen, Z.~Yan, and G.~Liang (2007), Toward an
  automated parallel computing environment for geosciences, \textit{Physics of
  the Earth and Planetary Interiors}, \textit{163}(1-4), 2--22.

\end{thebibliography}

\end{document}